\documentclass[%
 reprint,
 amsmath,amssymb,
 aps,
]{revtex4-2}

\usepackage{graphicx}
\usepackage{dcolumn}
\usepackage{bm}

\usepackage{xcolor}

\newcommand{\f}{\frac}

\newcommand{\de}{\delta}

\newcommand{\om}{\omega}
\newcommand{\Om}{\Omega}

\newcommand{\ta}{\theta}

\newcommand{\la}{\lambda}

\begin{document}

\preprint{APS/123-QED}

\title{Coalescence of limit cycles in the presence of noise}

\author{Sergei Shmakov}
 \email{sshmakov@uchicago.edu}
\author{Peter B. Littlewood}
 \email{littlewood@uchicago.edu}
\affiliation{James Franck Institute and Department of Physics, The University of Chicago, Chicago, Illinois 60637, United States
}

\date{\today}

\begin{abstract}
Complex dynamical systems may exhibit multiple steady states, including time-periodic limit cycles, where the final trajectory depends on initial conditions. With tuning of parameters, limit cycles can proliferate or merge at an exceptional point. Here we ask how dynamics in the vicinity of such a bifurcation are influenced by noise. A pitchfork bifurcation can be used to induce bifurcation behavior. We model a limit cycle with the normal form of the Hopf oscillator, couple it to the pitchfork, and investigate the resulting dynamical system in the presence of noise. We show that the generating functional for the averages of the dynamical variables factorizes between the pitchfork and the oscillator. The statistical properties of the pitchfork in the presence of noise in its various regimes are investigated and a scaling theory is developed for the correlation and response functions. The analysis is done by perturbative calculations as well as numerical means. Finally, observables illustrating the coupling of a system with a limit cycle to a pitchfork are discussed and the phase-phase correlations are shown to exhibit non-diffusive behavior with universal scaling.
\end{abstract}

\maketitle

\section{\label{sec:intro}Introduction}
Dynamical systems exhibiting bifurcations - where there is a sudden change of state - are of  importance in many fields of study, ranging from biological sciences to quantum systems \cite{Strogatzbook,Guckenheimerbook,Kuramotobook}. For example (relevant to the present work) the onset of polariton condensation and lasing corresponds to collective synchronization of oscillators \cite{Dicke1954,Rice1994,Eastham2001,Kirton2018}, and the development of superconductivity and superfluidity can be mathematically described in the same way, perhaps most easily understood in the context of exactly solvable mean field Richardson-Gaudin models \cite{Dukelsky2004}. In biology the proximity to an instability can be used to tune the sensitivity of a sensor to an external stimulus \cite{winfree2001}, seen in the tuning of hair bundles in the cochlea \cite{Camelet2000}. Many physiological functions depend on the maintenance of synchronised patterns in time (limit cycles), including breathing \cite{Smith1991}, cardiac function, and circadian rhythms \cite{Rijo-Ferreira2019}. Photonic oscillators such as optically injected lasers also exhibit limit cycles, allowing for unique tunability and response properties \cite{Himona2022, Vassilios2021}.  

Bifurcations in spatially extended dynamical systems may fall into generalized classes, and the onset of oscillatory behavior from a stationary state bears analogy to a phase transition in a thermodynamic system. There is a particular subclass of such transitions which is characterized by an exceptional point of the dynamics, where two modes merge with identical eigenvalues \cite{Hanai2019,Fruchart2021}. This critical point will be the focus of our paper.

All physical systems operate in the presence of environmental fluctuations and noise. This can mean that dynamical transitions in finite systems (or infinite systems below the lower critical dimension) are blurred. However, dynamical responses can still be tuned - the classic case is usually called stochastic resonance \cite{Mcnamara, SR2}, where a bistable system in the presence of noise is found to be maximally sensitive to a.c. fields whose frequency matches the typical rate of hopping between minima. As an example of a bistable system (at least in one regime), we take a pitchfork bifurcation, which can be coupled to more general dynamical systems in order to imbue those with bifurcation behavior. Unlike the pitchfork itself, these coupled systems can exhibit much broader classes of dynamics, where the coalescing states can be fixed points, various stable orbits such as limit cycles, and other attractors \cite{Weis}. In this paper we study such a dynamical system: noiseless dynamics can merge two limit cycles (by parameter tuning) at an exceptional point. What are the response and correlation functions of this system in the presence of noise? 

Noisy limit cycles and oscillators are of considerable interest in many fields of study, and have been explored in various contexts \cite{LimitCycleNoise,Example1,Example2,Example3}. In the vicinity of the bifurcation, particular dynamical systems can be expressed in their normal forms that are generic to a class of models. These normal forms can therefore be studied to infer general information on how coupled dynamical systems with bifurcations behave. In this paper we will investigate a Hopf oscillator coupled to the pitchfork, which can describe a class of systems with coalescing limit cycles. We will show that there are characteristic features in the dynamics that mix the established scaling of stochastic resonance with additional time scales associated with the recurrent limit cycle oscillations.

One reason to study critical points is that they may have universal dynamical signatures. The existence of a scaling theory for a model system (which we will derive in this case) should, near the critical point, provide a universality class for dynamical correlations of systems with the same underlying symmetry. Consequently experimental systems that are microscopically much more complicated than our model may nonetheless have universal features in their correlation functions which reveal underlying dynamical transitions. Our work will suggest methods to analyze data to explore whether the bifurcation model is appropriate for the system at hand.
 
The nature of the noise depends on the details of the system, and in this paper we investigate the simplest case: Gaussian white noise. One of the main results of this paper, which describes how expectation values of operators factorize between the pitchfork and the rest of the dynamical system, are valid for any type of noise, as long as it is uncorrelated between the pitchfork and the rest of the system. The behavior of the pitchfork with noise has been investigated in various contexts and in various regimes, such as perturbative methods \cite{Hertz} and stochastic resonance \cite{Mcnamara, SR2}. However, to authors' knowledge, there is no treatment that combines the noisy pitchfork in it's various regimes into one theory. Here we investigate and collect results for it in all  regimes, and sew them together with scaling theory. That theory relies on the observation that the pitchfork has a very limited number of dimensionless parameters that the perturbative expansions can run over. 

Hence, once the factorization result and the scaling theory of the pitchfork are explored, our goal is to identify observables that can be measured to see whether an experimental system at hand does indeed exhibit factorization and whether it exhibits a bifurcation governed by the pitchfork. Other possible venues of exploration for future study include investigating other types of orbits that can bifurcate coupled to a pitchfork, other types of noise, and also noise that is correlated between the pitchfork and the rest of the dynamical system.

The reason to investigate an abstract model is that it may offer universal pointers for physical behavior. Slow fluctuations between different conceptual states are common in complex systems and are often loosely assumed in biology. It is almost never the case that one will have a complete enough understanding of the system in order to produce a detailed model. We rely instead on general principles, which are necessarily mostly static. However, the statistics of fluctuations near to the critical point can be deterministic of the underlying model. We will show that fluctuations between limit cycles have a greater richness than fluctuations between fixed points (traditional stochastic resonance). Such correlations could be determined from data, agnostic from a model. 

This paper is organized as follows. In Sec. \ref{sec:model} we introduce the model system of a Hopf oscillator coupled to a pitchfork bifurcation and demonstrate that it can be factorised into noisy dynamics of the pitchfork and of the limit cycle, eqn. (\ref{eq:factorization}). Here we also explore what that result implies for various observables of the limit cycle, eqns. (\ref{eq:anglecorr}), (\ref{eq:angleresponse}), (\ref{eq:phasecorr}), (\ref{eq:diffusion}). In Sec. \ref{sec:scaling} we perform a scaling analysis of the noisy pitchfork (which is a classic model for stochastic resonance). In Sec. \ref{sec:numerics} we demonstrate the numerical results and discuss how physical correlation functions of variables that correlate both the periodic "limit cycle" space and the "pitchfork" space couple and lead to novel interference features in the dynamics. In Sec. \ref{sec:conclusion} we close with a discussion, and in particular talk about  potential experimental impacts.\\

\section{\label{sec:model}Theoretical Model}
We take a Hopf oscillator as a model for a limit cycle and couple it to a pitchfork bifurcation:
\begin{eqnarray}
    &\dot{\psi} = (a+i\Om+b|\psi|^2)\psi+ic\psi w^n \;,\\
    &\dot{w} = Rw-gw^3 \;.
\end{eqnarray}
Here $a,\Om,b$ are real parameters chosen to produce a limit cycle, $c$ is the coupling strength between the pitchfork and the oscillator and $R$ and $g>0$ are parameters of the pitchfork. All parameters are of appropriate units and are taken to be constant in time. We assume that the Hopf oscillator is in a regime with a stable limit cycle ($a>0, b<0$), although the particular values of the parameters will not be relevant for most of the analysis. This system has been shown to exhibit bifurcation behaviour and for $n=1$ and $n=2$ there is also an exceptional point at $R=0$ \cite{Weis}. Notice that in this case the Hopf oscillator is coupled to the pitchfork in a multiplicative way. Since $\psi$ is complex, we can break it up into its real and imaginary components. We then add noise with the assumption that it is not correlated between the the pitchfork and the oscillator.
\begin{eqnarray*}
    &\dot{\psi}_x = a\psi_x-\Om\psi_y+b|\psi|^2\psi_x-c\psi_y w^n+\sqrt{B_{\psi_x}(\psi)}\xi_{1}(t)\;,\\
    &\dot{\psi}_y = a\psi_y+\Om\psi_x+b|\psi|^2\psi_y+c\psi_x w^n+\sqrt{B_{\psi_y}(\psi)}\xi_{2}(t)\;, \\
    &\dot{w} = Rw-gw^3 + \sqrt{B_{w}(w)}\xi_{3}(t) \;,
\end{eqnarray*}
where the stochastic differential equations (SDEs) will be taken under Ito discretization and the noise increments are defined to be independent Gaussian white noises $\left<\xi_i(t_1)\xi_j(t_2)\right> = \de_{ij}\de(t_1-t_2)$. We will assume that the noises along the orthogonal directions of the Hopf oscillator are equal: $B_{\psi_x}=B_{\psi_y}=B_\psi$. Calling $\psi_x = r\cos(\ta)$ and $\psi_y = r\sin(\ta)$ we can convert the Hopf oscillator into polar coordinates, keeping the terms produced due to Ito discretization:
\begin{eqnarray}
    &\dot{r} = ar+br^3+\f{B_\psi}{r}+\sqrt{B_\psi}\xi_r(t) \;, \\
    &\dot{\ta} = \Om+cw^n + \f{\sqrt{B_\psi}}{r}\xi_{\ta}(t) \;,
\end{eqnarray}
where $\xi_r(t)$ and $\xi_\ta(t)$ are defined through an orthogonal transformation from the original $\xi_1(t)$ and $\xi_2(t)$, and are also Gaussian white uncorrelated noises \cite{Gardiner}. We can see here that the coupling to the pitchfork only affects the dynamics of $\ta$, as in the corresponding classical case, and in fact the dynamics of $r$ are completely independent of both $\ta$ and $w$. 

We can now introduce the Martin-Siggia-Rose response fields $w^q$, $r^q$ and $\ta^q$ and apply the Janssen-De Dominicis-Peliti formalism \cite{MSR1,MSR2,Atland,Kamenev} to express the problem in terms of path integrals:
\begin{widetext}
\begin{equation}
\begin{split}
    \mathcal{Z} =  \int \mathcal{D}[w,w^q,\ta,\ta^q,r,r^q]  &\exp \left[\int dt\, iw^q(\dot{w}-Rw+gw^3)-\f{B_w}{2}w^{q2}\right] \exp\left[\int dt\, i\ta^q(\dot{\ta}-\Om-cw^n)-\f{B_\psi}{2r^2}\ta^{q2}\right]\\&*\exp\left[\int\, dt ir^q(\dot{r}-f(r))-\f{B_\psi}{2}r^{q2}\right] \;,
\end{split}
\end{equation}
\end{widetext}
where $\mathcal{Z} = 1$ identically, but can be used as a generating functional for operators comprised of our dynamic variables $r,\ta,w$. Here $f(r)=ar+br^3+B_{\psi}/r$ contains the deterministic part of the evolution of $r$. We will assume unspecified initial conditions, in other words the systems were settled into their equilibrium distributions at $t = -\infty$. Here we also note that we have not taken into account the compactness of $\ta$, which in principle should be handled. However, the following result still holds and is confirmed numerically.  We can now demonstrate that expectation values of these operators factorize into expectation values of the constituents corresponding to field theories of the Hopf oscillator and pitchfork on their own. That fact relies on the linearity of the equation for $\ta$. It can be clearly seen that when calculating an expectation value of any operator $w^n(w^{q})^m$, we can integrate out the fields corresponding to the Hopf oscillator and we'll just be left with the dynamics of the pitchfork. Therefore, the complexity comes from any parts of the operators of the form $\ta^n(\ta^{q})^mr^{k}(r^{q})^l$. These operators can be obtained from a generating functional:
\begin{equation}
\begin{split}
    &\mathcal{Z}[\la_\ta,\la^q_\ta,\la_r,\la^q_r]= \\& \left<\exp\left[\int dt \la_\ta\ta+\la_\ta^{q}\ta^q +\la_rr+\la_r^{q}r^q \right]\right> \;,   
\end{split}
\end{equation}
where all $\la$'s are functions of time. Employing the linearity of the equation for $\ta$ it can be shown by integrating out $\ta$ and $\ta^q$ that this generating functional factorizes into the generating functional for the Hopf oscillator in the absence of the coupling and an expectation value of an appropriate operator of the pitchfork only:
\begin{equation}
\begin{split}
    &\mathcal{Z}[\la_\ta,\la^q_\ta,\la_r,\la^q_r]=\mathcal{Z}_{\text{Hopf}}[\la_\ta,\la^q_\ta,\la_r,\la^q_r]*\\&\left<\exp\left[\int dt \, cw^n\int_t^{\infty}ds\, \la_\ta(s)\right]\right>_w   \;,
\end{split}
\label{eq:factorization}
\end{equation}
where $\left<...\right>_w$ is the expectation value with respect to the pitchfork action:
\begin{equation}
    S_w[w,w^q]=\int dt\,\left(iw^q(\dot{w}-Rw+gw^3)-\f{B_w}{2}w^{q2}\right) \;.
\end{equation}

Hopf oscillators in their normal form have been studied in many contexts, with varying mathematical depths and in application to a variety of subjects, both with and without noise \cite{HopfExample1, HopfExample2, HopfExample3, HopfExample4, HearingHopf}. Thus, we won't address part of the generating functional $Z_{Hopf}$ that comes from the dynamics of the Hopf oscillator alone. However, we can look at some quantities of interest of the model and see how the coupling to the pitchfork affects them. For example, for $n=1$ the connected angle correlation function for the full system can be shown to be:
\begin{equation}
    \begin{split}
        &\left<\ta(t_1)\ta(t_2)\right>_c = \f{\de}{\de \la_\ta(t_1)}\f{\de}{\de \la_\ta(t_2)}\ln \mathcal{Z}[\la_\ta,\la^q_\ta,\la_r,\la^q_r] \bigg|_{\la=0} \\&=\left<\ta(t_1)\ta(t_2)\right>_{c,Hopf}+c^2\int^{t_1}dt\int^{t_2}dt'\left<w(t)w(t')\right>_{c,w} \;.
    \end{split}
    \label{eq:anglecorr}
\end{equation}

On the other hand, the phase response function of the coupled system is unchanged by the presence of the pitchfork:
\begin{equation}
    \begin{split}
        &\left<\ta(t_1)\ta^q(t_2)\right>_c = \f{\de}{\de \la_\ta(t_1)}\f{\de}{\de \la^q_\ta(t_2)}\ln \mathcal{Z}[\la_\ta,\la^q_\ta,\la_r,\la^q_r] \bigg|_{\la=0} \\&=\left<\ta(t_1)\ta^q(t_2)\right>_{c,Hopf}\;.
    \end{split}
    \label{eq:angleresponse}
\end{equation}

More importantly, since the object of interest is a limit cycle, the observable of interest will be phase-phase correlations, which, due of the factorization discussed above, will take the form:
\begin{equation}
    \begin{split}
        &\left<\exp\left[i\ta(t_1)-i\ta(t_2)\right]\right> =\\& \left<\exp\left[i\ta(t_1)-i\ta(t_2)\right]\right>_{Hopf}\left<\exp\left[-ic\int_{t_1}^{t_2} dt w^n(t)\right]\right>_{w}\;.
    \end{split}
    \label{eq:phasecorr}
\end{equation}

Another object of interest might be the phase diffusion constant \cite{LimitCycleNoise}, which we will consider in the absence of external fields, leaving it time-translation invariant:
\begin{equation}
    \begin{split}
        D &\equiv \int_{-\infty}^{\infty} d\tau \left<\left(\dot{\ta}(t)-\left<\dot{\ta}\right>(t)\right)\left(\dot{\ta}(t+\tau)-\left<\dot{\ta}\right>(t+\tau)\right)\right>\\&=\int_{-\infty}^{\infty} d\tau B_{\psi}\left<\f{\de(\tau)}{r(t)r(t+\tau)}\right>\\&+c^2\left(\left<w^n(t)w^n(t+\tau)\right>-\left<w^n(t)\right>\left<w^n(t+\tau)\right>\right) \; ,
    \end{split}
    \label{eq:diffusion}
\end{equation}

where we used the independence of statistics of $\xi_\ta$, $w$ and $r$ to arrive at the second equality. For $n=1$, the diffusion constant is directly dependent of the connected two-point function of the pitchfork. For $n=2$ it will depend on the variance of the pitchfork distribution (which is just the two-point function at zero time separation) as well as a particular form of the four-point function. We can therefore see that to answer general questions about the coupled system, it is useful to understand the dynamics of the stochastic pitchfork bifurcation on it's own, and it's correlation functions.

\section{\label{sec:scaling}Pitchfork Scaling Theory}

The noisy, forced pitchfork obeys the stochastic differential equation (SDE):
\begin{equation}
    \dot{w} = Rw-gw^3+h(t)+\sqrt{B_w}\xi_3(t) \;,
\end{equation}
and the bifurcation occurs at $R=0$ in the classical system. It's not possible to obtain the correlation and response functions for the pitchfork in all regimes exactly, so we instead employ a scaling theory and patch together different regimes. In the SDE, we assume that $g>0$ and $B_w\geq0$. The parameters and the functions of interest all have dimensions restricted by the dimensions of $w$. We can therefore define:
\begin{widetext}
\begin{subequations}
\begin{align}
    \left<w\right> &\equiv M(R,g,B_w,h) = \left(\f{B_w}{g}\right)^{\f{1}{4}} F_1\left(\f{R}{B_w^{\f{1}{2}}g^{\f{1}{2}}}, \f{h}{B_w^{\f{3}{4}}g^{\f{1}{4}}}\right)\;,\\
    \left<w(\Delta t)w(0)\right>_c &\equiv C(R,g,B_w,h,\Delta t) = \left(\f{B_w}{g}\right)^{\f{1}{2}}F_2\left(\f{R}{B_w^{\f{1}{2}}g^{\f{1}{2}}},\f{h}{B_w^{\f{3}{4}}g^{\f{1}{4}}},g^{\f{1}{2}}B_w^{\f{1}{2}}\Delta t\right)\;,\\
    \left<w(\Delta t)w^q(0)\right>_c &\equiv \chi(R,g,B_w,h,\Delta t) =i\Theta(\Delta t) F_3\left(\f{R}{B_w^{\f{1}{2}}g^{\f{1}{2}}},\f{h}{B_w^{\f{3}{4}}g^{\f{1}{4}}},g^{\f{1}{2}}B_w^{\f{1}{2}}\Delta t\right)\;,
\end{align}
\end{subequations}
\end{widetext}
where $F_1,F_2,F_3$ are dimensionless functions of dimensionless variables. Similar scaling theories have been defined for the pitchfork, especially in the context of stochastic resonance \cite{SR2}, but here we proceed to define it for completeness. By $\left<...\right>_c$ we mean the connected correlation function, or cumulant. Several assumptions were made. First, we assume unconstrained initial conditions, so the system is time-translation invariant (at least in the absence of forcing). Second, all other dimensionless parameters that might arise from parameters implicit in the definition of $h(t)$ are present in these functions, but are not explicitly written for brevity. Third, causality is assumed for the response function $\chi$ \cite{Hertz, Chow}, and the Heaviside function $\Theta(t)$ is defined in the usual manner with the prescription that $\Theta(0)=0$ to be consistent with Ito discretization \cite{Chow}.

These functions are universal for any collection of parameters. Suppose some crossover is found that is intrinsic to these functions, for example in $F_2(x,0,z)$ for some $x$. The universality of these functions then guarantees that in physical parameters this crossover or transition has to behave as $R \sim \sqrt{gB_w}$, with the exact proportionality constant to be determined. The only way for this relationship to be broken is through the prefactors that give these functions proper dimensions to match the physical quantities of interest. However, these prefactors are also defined universally, and thus their presence can be integrated generically into understanding of such crossovers.

We can now extract general information about the functions $F_1,F_2,F_3$. No attempt was made to study these functions for all values of the parameters of the system. However, various regimes allow for use of different techniques to compute these functions. We define the effective bifurcation parameter $\tilde{R} =R/\sqrt{gB_w}$. For very large and negative $\tilde{R}$, it is possible to apply the usual field-theoretical perturbative techniques \cite{Chow, Hertz, Barabasi,Kamenev,Atland,Binney}. For large and positive $\tilde{R}$, the system is going to hop between the two wells so that we cannot just expand the field theory around one of the wells, but rarely enough so that the system has enough time to settle into either well before hopping again. In other words, the relaxation around either stable state will be negligible for long-term dynamics. Here it is possible to approach the study of the functions of interest from the perspective of Kramer's problem and stochastic resonance \cite{Kamenev,Mcnamara}. There will be two relevant time scales. The first one is responsible for the decay within either well, and can be obtained by expanding the action around either well. The lowest order approximation of this time scale is $\left(2R\right)^{-1}$. The other one is responsible for the rate of hopping between the wells and can be obtained through standard Kramers problem calculation to be $\tau \approx \sqrt{2}\pi R^{-1}\exp\left[R^2/(2gB_w)\right]$. The crossover between the two is therefore governed by the competition of the two time scales. Unsurprisingly, perhaps, the ratio of these time scales is also a function of the effective bifurcation parameter: $\pi/\sqrt{2}\exp\left(\tilde{R}^2/2\right)$. The final regime centered around $R=0$ is mainly studied numerically. We also define the effective time $\tilde{t}=\sqrt{gB_{w}}t$ and the effective external field $\tilde{h}=h/(B_{w}^{\f{3}{4}}g^{\f{1}{4}})$.

\section{\label{sec:numerics}Numerical results}
\subsection{\label{sec:pitchforkNum}Pitchfork}
\begin{figure}[!htb]
\includegraphics[scale = 0.18]{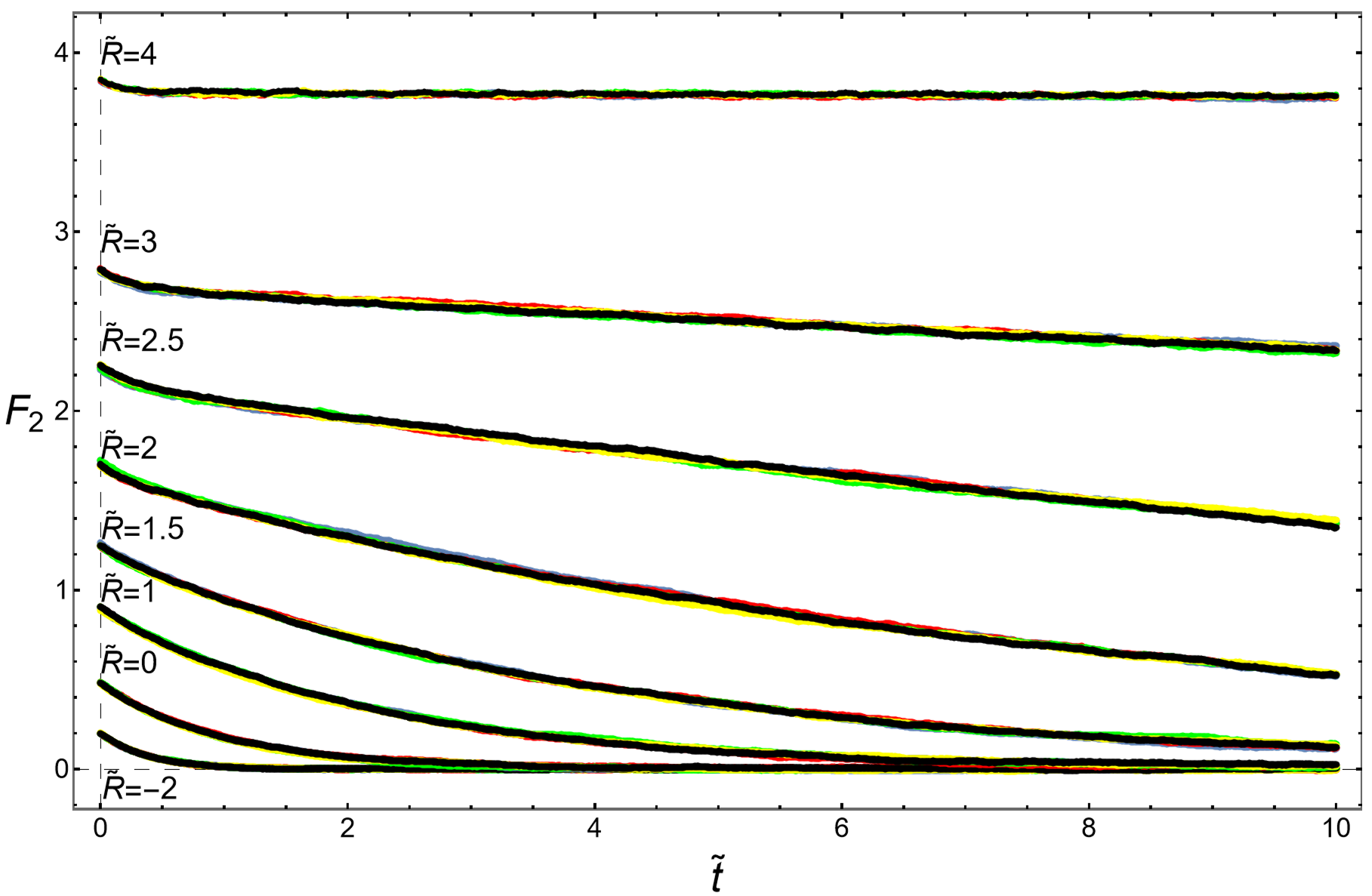}
\caption{\label{fig:scaling overlap} Numerical simulation of $F_2(\tilde{R},0,\tilde{t})$ as a function of $\tilde{t}$ at various $\tilde{R}$. Each line is an overlap of 5 simulations, in different colors, with different values of the individual parameters that leave $\tilde{R}$ the same.}
\end{figure}

To demonstrate and support the results of the scaling theory, we turn to numerical simulations of our random process. First, we consider the pitchfork itself. In Fig.~\ref{fig:scaling overlap}, we simulate the process and plot the non-dimensionalized correlation function $F_2(\tilde{R},0,\tilde{t})$ for different values of $\tilde{R}$. Each line on the plot is an overlap of five results of simulations (in different colors) at different values of $R$, $B_w$ and $g$ that leave $\tilde{R}$ invariant. Their complete overlap demonstrates the validity of the developed scaling theory.

\begin{figure}[!htb]
\includegraphics[scale = 0.18]{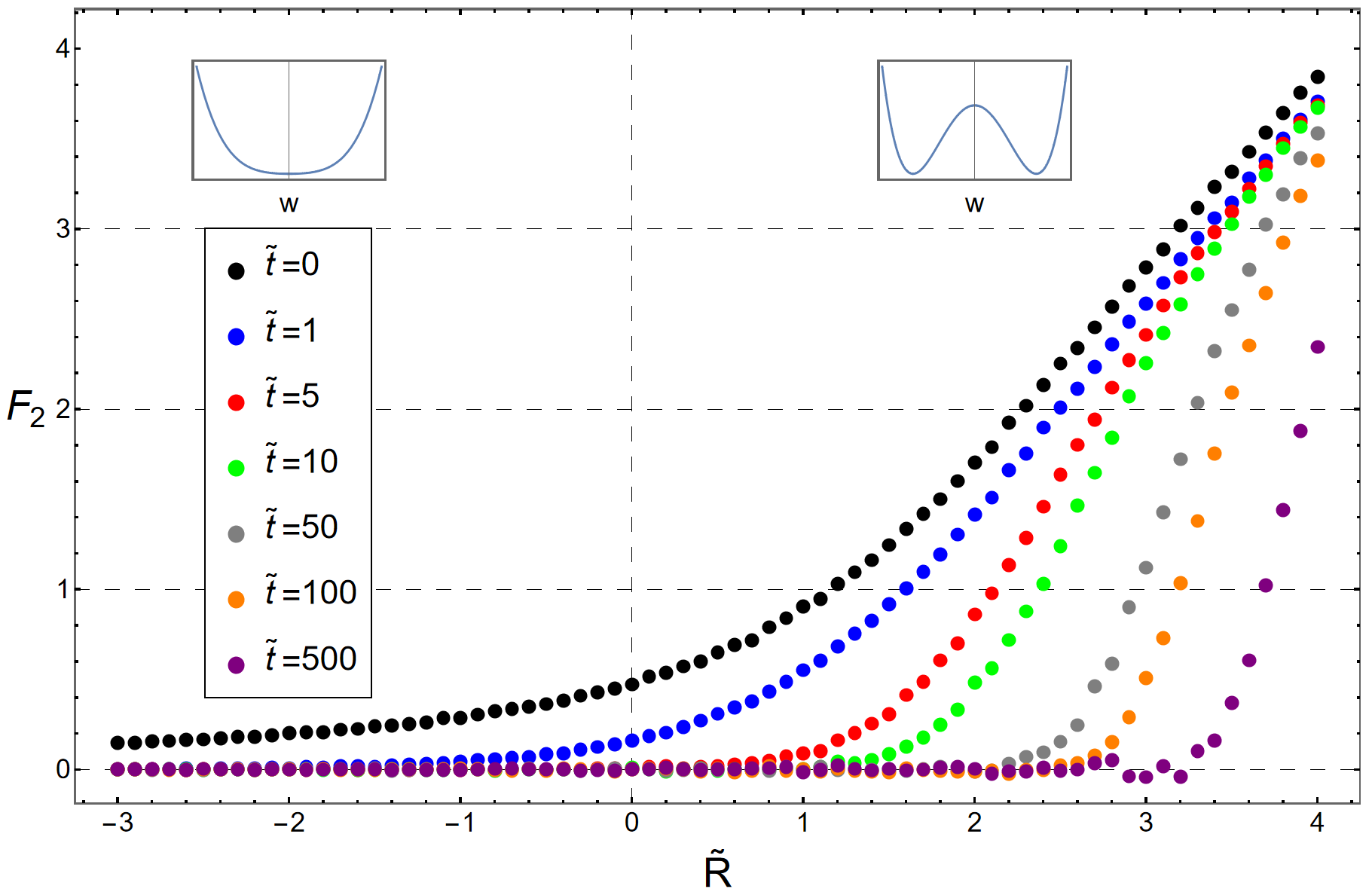}
\caption{\label{fig:time slices} Numerical simulation of $F_2(\tilde{R},0,\tilde{t})$ as a function of $\tilde{R}$ at various $\tilde{t}$.}
\end{figure}

We expect that the decay of the correlation function will be governed by time-scales that are described above. In Fig.~\ref{fig:time slices} we plot the time slices of $F_2(\tilde{R},0,\tilde{t})$ as a function of $\tilde{R}$ for different values of $\tilde{t}$. As can be seen, there is a smaller time scale associated with the immediate decay of the correlation function, which perturbatively we expect to be $~R^{-1}$ in the $\tilde{R}<<0$ regime and $~(2R)^{-1}$ in the $\tilde{R}>>0$ regime. In the $R\leq0$ regime that is the only time scale. In $R>0$ regime there is also Kramers time scale which is exponential and controls the decay starting with values of smaller and smaller $\tilde{t}$ as $\tilde{R}$ increases.

\subsection{\label{sec:phase}Phase-phase correlations}
\begin{figure}[!htb]
\includegraphics[scale = 0.18]{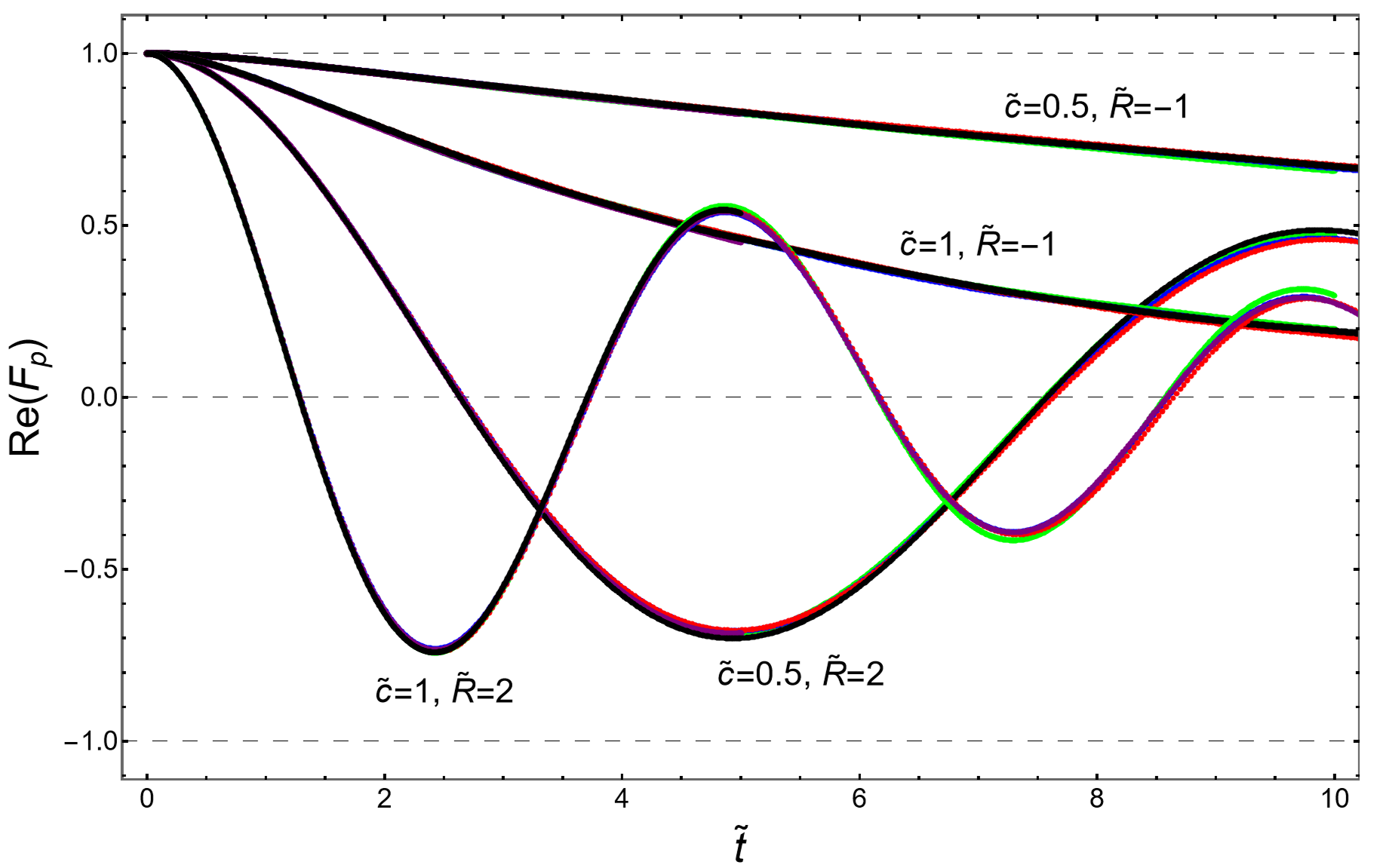}
\caption{\label{fig:n1 scaling} Numerical simulation of $Re[F_p(\tilde{R},0,\tilde{t},\tilde{c})]$ for $n=1$ as a function of $\tilde{t}$ at various $\tilde{R}$ and $\tilde{c}$. Each line is an overlap of 5 simulations, in different colors, with different values of individual parameters that leave the dimensionless variables the same.}
\end{figure}

\begin{figure}[!htb]
\includegraphics[scale = 0.18]{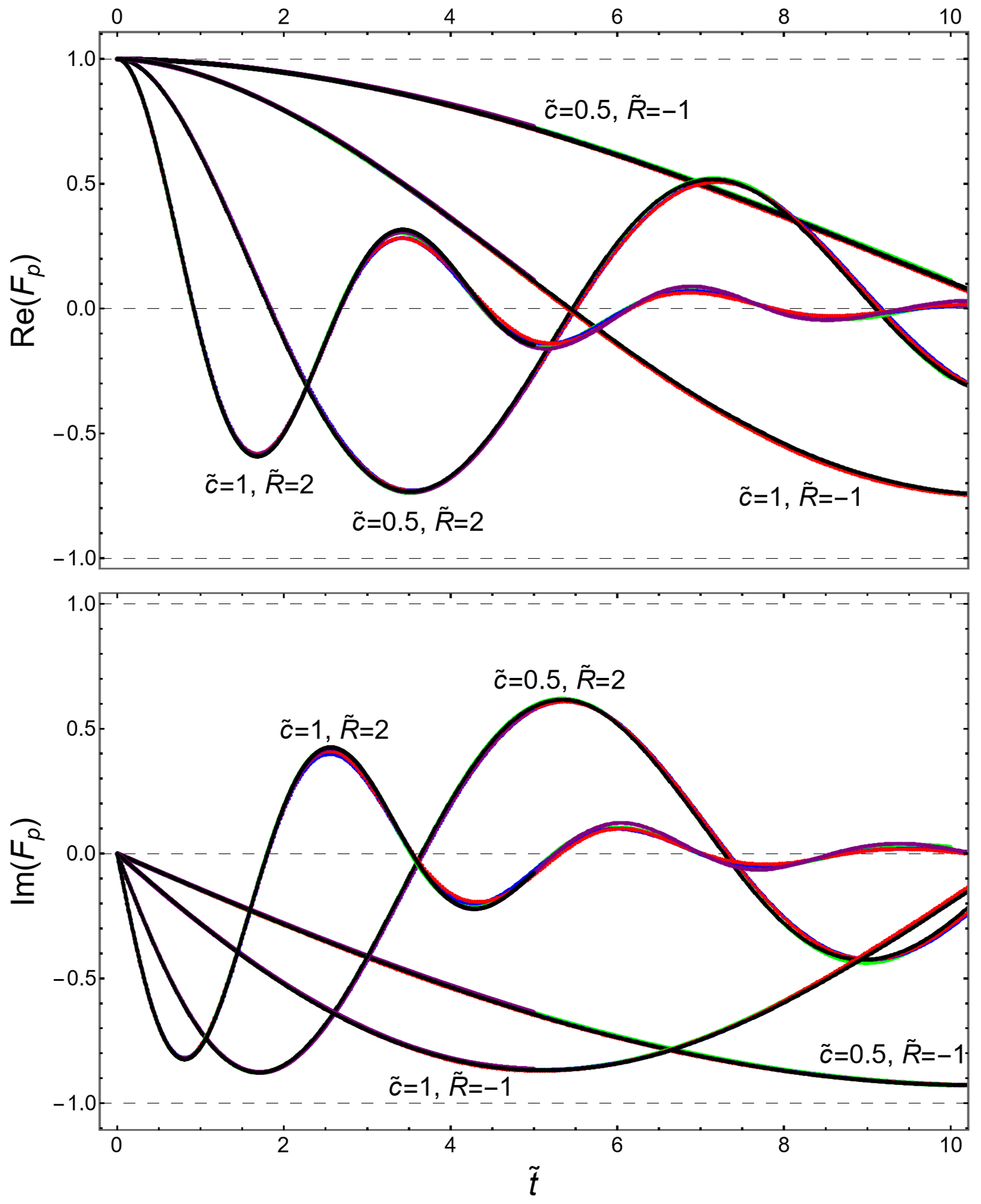}
\caption{\label{fig:n2 scaling} Numerical simulation of $Re[F_p(\tilde{R},0,\tilde{t},\tilde{c})]$ and $Im[F_p(\tilde{R},0,\tilde{t},\tilde{c})]$ for $n=2$ as a function of $\tilde{t}$ at various $\tilde{R}$ and $\tilde{c}$. Each line is an overlap of 5 simulations, in different colors, with different values of individual parameters that leave the dimensionless variables the same.}
\end{figure}

Now we can return back to the original system. Since the information about the period and the diffusion due to noise of the limit cycle is contained in the phase-phase correlations, the operator of interest is the one in Eq.~(\ref{eq:phasecorr}). This operator can also be subjected to scaling theory, and with an additional parameter - the coupling constant $c$, there is an additional dimensionless parameter: $\tilde{c}=cB_w^{\f{n}{4}-\f{1}{2}}g^{-\f{n}{4}-\f{1}{2}}$. The phase correlations factor coming from the pitchfork is dimensionless, therefore we call this factor $F_p(\tilde{R},\tilde{h},t_{2,eff}-t_{1,eff},\tilde{c})$. The scaling is demonstrated in Fig.~\ref{fig:n1 scaling} for $n=1$ and in Fig.~\ref{fig:n2 scaling} for $n=2$, where we plot $F_p$ for various values of $\tilde{R}$ and $\tilde{c}$ as a function of $\tilde{t}$ (and there is no external field).  With this, the two cases of $n=1$ and $n=2$ can be studied separately.

\subsection{\label{sec:n1}$n=1$}
First, by formal expansion it can be seen that for $n=1$, all the imaginary terms will have an odd power of $w$. In the absence of external fields the pitchfork SDE has a $Z_2$ symmetry, and thus they will all be zero identically, and only the real part of the phase-phase correlations will survive. To formally see that the dimensionless parameter above (which is $cB_w^{-\f{1}{4}}g^{-\f{3}{4}}$ for $n=1$) is indeed what controls the expansion, we can employ the scaling theory to describe the lowest order terms:
\begin{equation}
\begin{split}
    &F_p = \left<\exp\left[-ic\int_{t_1}^{t_2} dt w(t)\right]\right>_{w}\approx1- \\& \f{c^2}{2g^{\f{3}{2}}B_w^{\f{1}{2}}}\int_0^{\tilde{t}_2-\tilde{t}_1}dz'\int_{-z'}^{\tilde{t}_2-\tilde{t}_1-z'}dzF_2(\tilde{R},\tilde{h},z)    \; . 
\end{split}
\end{equation}

As an example, the following limit of this function can be calculated:
\begin{equation}
    \lim_{\tilde{R}\rightarrow -\infty} F_p(\tilde{R},0,\tilde{t},\tilde{c}) = \exp \left[-\tilde{c}^2\left(\f{\tilde{t}}{2\tilde{R}^2}-\f{e^{\tilde{R}\tilde{t}}}{2\tilde{R}^3}+\f{1}{2\tilde{R}^3}\right)\right] \; .
\end{equation}

The leading corrections to $\tilde{R}$ can be calculated perturbatively \cite{Hertz}.

For $n=1$ the contribution of the pitchfork to the diffusion constant is also conceptually quite simple:
\begin{equation}
\begin{split}
    D_{p,n=1} &= c^2\int_{-\infty}^{\infty} d\tau \left(\left<w(t)w(t+\tau)\right>-\left<w(t)\right>\left<w(t+\tau)\right>\right) \\&= \f{c^2}{g}\int_{-\infty}^{\infty} dz\, F_2(\tilde{R},\tilde{h},z) \; ,
\end{split}\;,
\end{equation}

\begin{figure}[!htb]
\includegraphics[scale = 0.18]{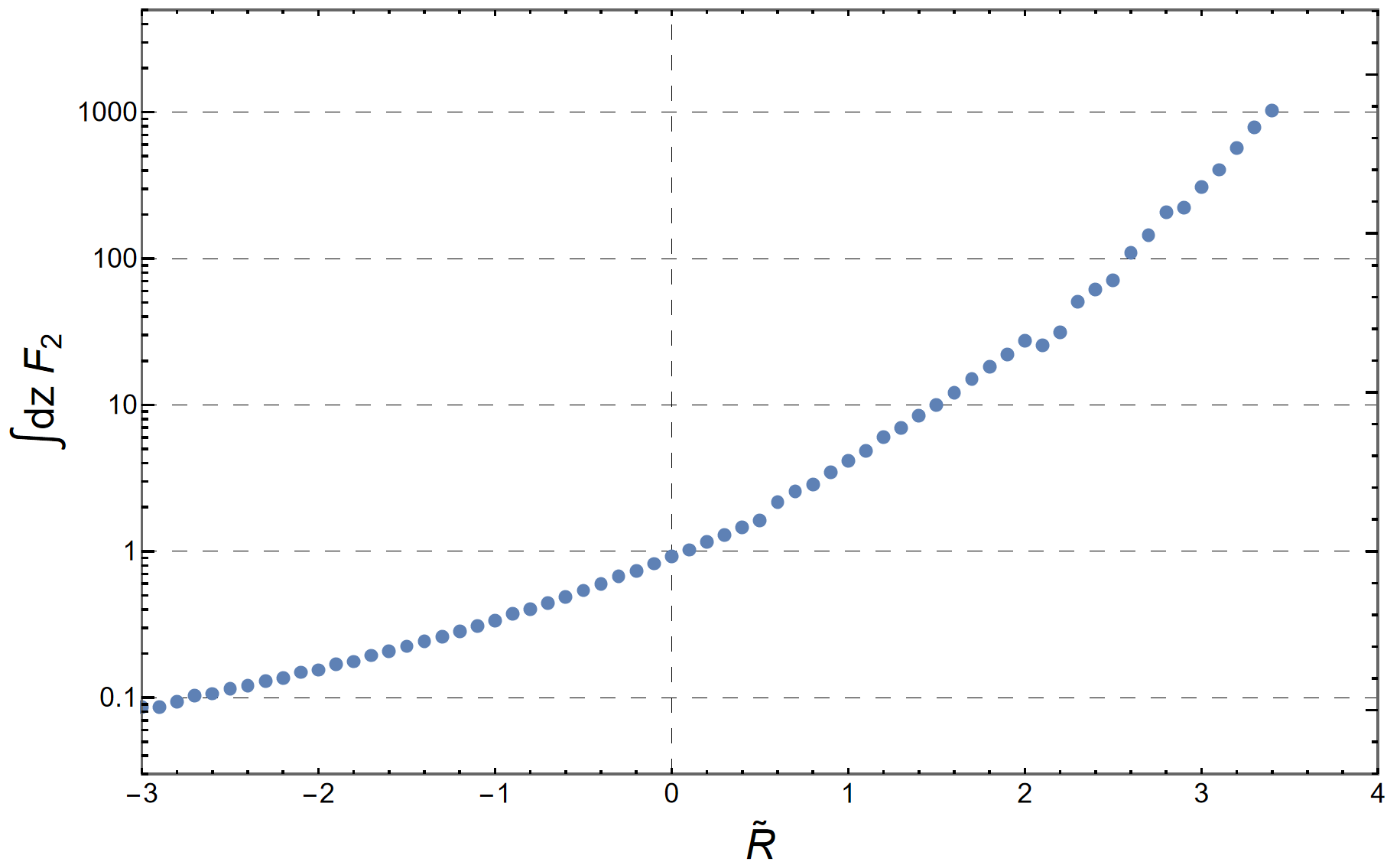}
\caption{\label{fig:diffusion} Numerical simulation of $\int_{-\infty}^{\infty}dz F_2$ as a function of $\tilde{R}$ on a logarithmic scale at $\tilde{h}=0$.}
\end{figure}

Where we used the definition of the dimensionless correlation function $F_2$ and non-dimensionalized the integral measure $z=\sqrt{gB_w}\tau$. Although we assumed time translation invariance, this expression demonstrates the scaling of the diffusion constant in the presence of forcing as well. The numerical results for the integral are demonstrated in Fig.~\ref{fig:diffusion} for various values of $\tilde{R}$. 

\subsection{\label{sec:n2}$n=2$}
The $n=2$ case is simpler conceptually. The coupling to the pitchfork changes the period of the limit cycle governed, to lowest order, by the variance of the pitchfork distribution. Since it is always positive, there are always oscillations in the phase-phase correlations, as well as modulation of the period. The oscillations can be understood intuitively: whatever the state of the pitchfork, it always drives the limit cycle along the positive phase direction, so at some points it is likely that on average the phase will be opposite to the starting phase, hence the correlations get a negative real part.

The effective coupling constant for the $n=2$ case is $\tilde{c}=cg^{-1}$, which is independent of pitchfork noise. Therefore, changing the noise strength of the pitchfork amounts to changing $\tilde{R}$ and thus, with corresponding tuning of $R$, can be scaled away (with the time axis being stretched as well).

\subsection{\label{sec:freq}Frequency space}
A useful representation of the phase-phase correlations that is experimentally viable given the access to the oscillator and demonstrates the crossover is the Fourier representation. In the absence of external fields acting on the pitchfork (so that the time translation invariance is preserved), the Fourier transform of phase-phase correlations takes the form:
\begin{equation}
\begin{split}
    &\int dt \left<\exp\left[-ic\int_{0}^{t} dt w^n(t)\right]\right>_{w}\exp(-i\om t)\\
    &=\int dt \left<\exp\left[-ic\int_{0}^{t} dt \left(w^n(t)+\f{\om}{c}\right)\right]\right>_{w} \;.
\end{split}
\end{equation}

\begin{figure}[!htb]
\includegraphics[scale = 0.18]{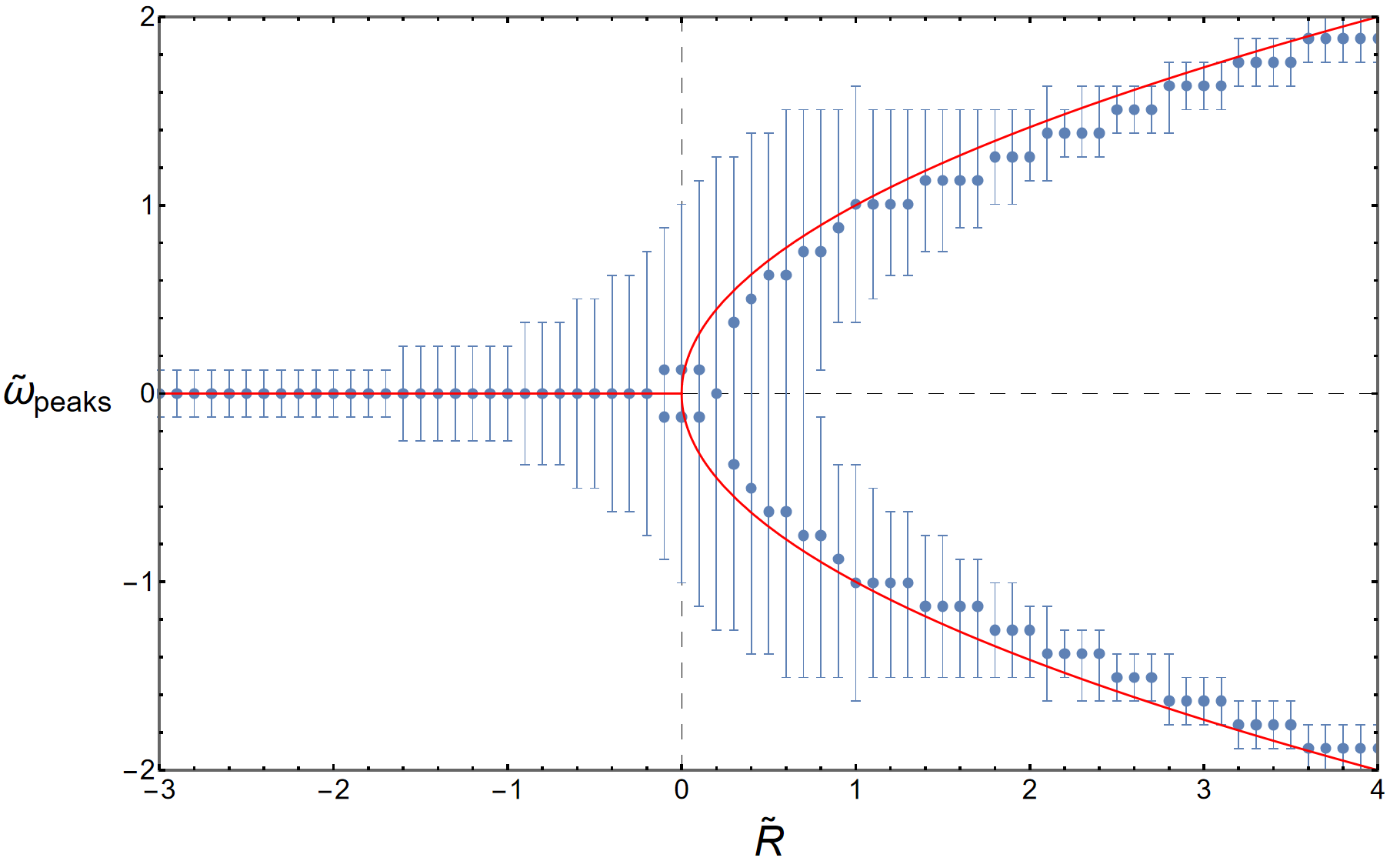}
\caption{\label{fig:n1peaklocations} Peak frequencies (blue dots) of the absolute value of the Fourier transform of the phase-phase correlations as a function of $\tilde{R}$ for $n=1$ and $\tilde{c}=1$. The bars represent the widths of half maxima of the peaks. In red are the mean field stable states of the pitchfork. The frequencies take discrete values due to the finite number of time steps in the simulations of the Ito processes. The frequency is offset by the natural frequency $\Om$ of the oscillator.}
\end{figure}

\begin{figure*}[!t]
\includegraphics[scale=0.24]{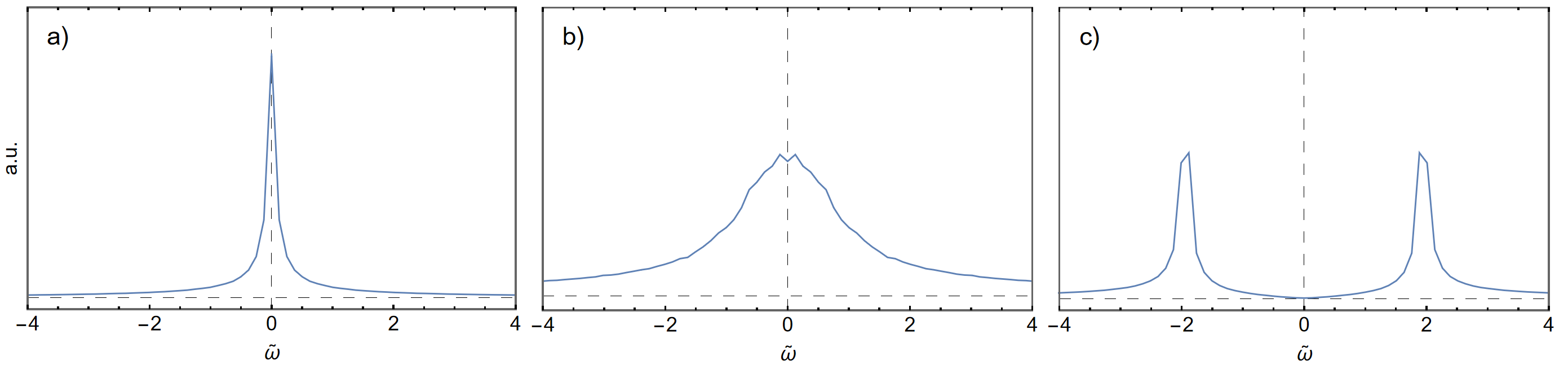}
\caption{\label{fig:fexamples} Example of absolute values of Fourier transforms of the phase-phase correlations for $n=1$ and $\tilde{c}=1$ versus dimensionless frequency $\tilde{\om} = \om/\sqrt{gB_w}$. The vertical axis is rescaled for each plot, and is thus measured in arbitrary units (a.u.). a) $\tilde{R}=-3$, b) $\tilde{R}=0$, c) $\tilde{R}=4$}
\end{figure*}

 An example of Fourier transforms for $n=1$ is given in Fig.~\ref{fig:fexamples}. The bifurcation in the pitchfork is clearly exhibited by the phase-phase correlations in frequency space. It is clear that the dominant contributions to the transform will come from the most likely states that the pitchfork can be found in. To demonstrate it further, we plot in Fig.~\ref{fig:n1peaklocations} the frequencies of the peaks of the Fourier distribution as a function of $\tilde{R}$, and see that they have the same bifurcation diagram as the pitchfork.

For $n=1$, the mean field stable states of the pitchfork for $R>0$ are $w=\pm\sqrt{R/g}$. Non-dimensionalizing the frequency of the Fourier transform as $\tilde{\om}=\om/\sqrt{gB_w}$, we get the scaling for the peak locations for $R>0$: $\tilde{\om}_{peaks}=\pm \tilde{c}\sqrt{\tilde{R}}$. 

For the case $n=2$, the expression for the mean field positions of the peaks takes the form $\tilde{\om}_{peaks} = -\tilde{c}\tilde{R}$, with the replacement of $\tilde{c}$ by the corresponding definition at $n=2$ (see above).
\begin{figure}[!htb]
\includegraphics[scale = 0.18]{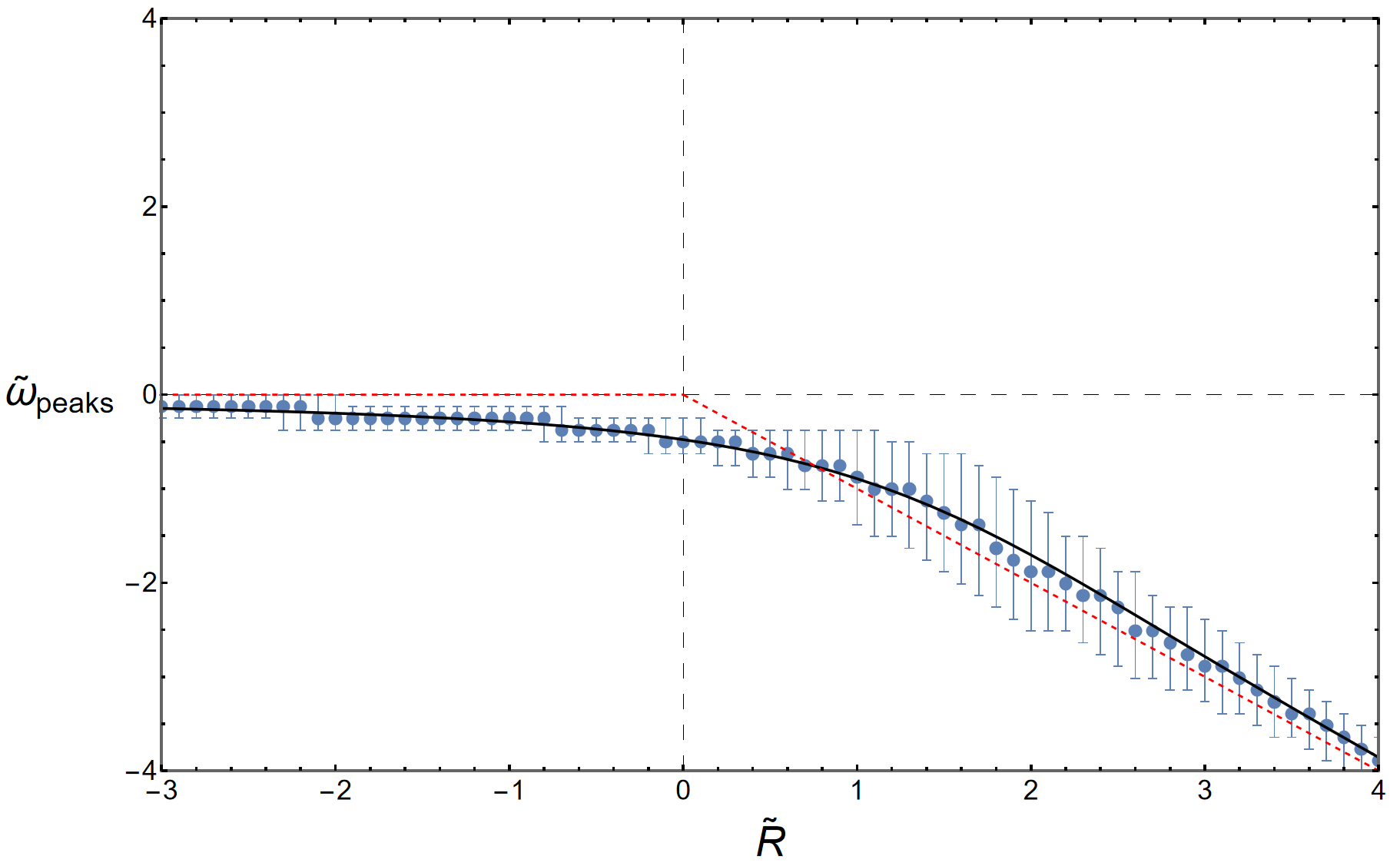}
\caption{\label{fig:n2peaklocations} Peak frequencies (blue dots) of the absolute value of the Fourier transform of the phase-phase correlations as a function of $\tilde{R}$ for $n=2$ and $\tilde{c}=1$. The bars represent the widths of half maxima of the peaks. The dashed red line are the mean field stable states of the pitchfork. The black line is $-\tilde{c}F_2(\tilde{R},0,0)$. As before, the frequencies take discrete values due to the finite number of time steps in the simulations of the Ito processes. The frequency is offset by the natural frequency $\Om$ of the oscillator.} 
\end{figure}
Notice that the peaks will only appear in the negative frequencies. However, the dominant contribution to the Fourier transform will come from the factor $\left<w(t)^2\right>$, which means the peak locations will approximately obey $\tilde{\om}_{peaks}=-\tilde{c}F_2(\tilde{R},0,0)$. These relationships are demonstrated in Fig. ~\ref{fig:n2peaklocations}.

The crossover in this case is not as clear as in the $n=1$ case. However, in both cases the widths of the peaks undergo a broadening around $R=0$ and are controlled by $\tilde{c}$. This can be used as an indication of a crossover.

As a final remark, the phase-phase correlations also have a term coming from the Hopf oscillator itself that would go into the Fourier transform. However, in the limit cycle regime, the dynamics of the oscillator are simple, in that the Fourier transform of the phase-phase oscillations has a peak at $-\Om$, the width of which depends on the noise strength. For the coupled system, it means that in the above discussion the zero frequency will be shifted to $-\Om$, and the peaks widths will in reality be broader. However, their locations and widths will undergo the same behavior described above, with the same scaling laws. 

\section{\label{sec:conclusion}Conclusion}
We have taken a model of bifurcating limit cycles that we believe to be generally applicable to a wide variety of systems near a bifurcation and explored it in the presence of noise. A pitchfork bifurcation is used to imbue a limit cycle with a bifurcating behaviour and the main result of the study is that this model system can be factorized into corresponding noisy dynamics of the pitchfork and the oscillator. 

The factorization of the generating functional depends only on the statistical independence between the noise acting on the pitchfork and the noise acting on the limit cycle. Therefore, specifics of the noise spectrum of either part of the system only determine the details of their individual effects. For example, a temporally correlated noise added to the pitchfork will add an additional dimensionless time scale to the scaling theory, therefore changing the above analysis. However, as the noise strength decreases, the scaling functions are driven further into their asymptotic regimes, where their behavior is easier to obtain analytically, and is also believed to be less sensitive to changes in the nature of noise and to presence of external fields. 

The effect of the pitchfork with Gaussian white noise on the oscillator was then explored, and a scaling theory was applied to demonstrate how the parameters of the pitchfork must come into the functions of interest of the limit cycle. The scaling theory is supported numerically, and measurable effects of the pitchfork on the limit cycle, such as on the phase diffusion constant and on the spectral resolution of the phase-phase correlations, are demonstrated.

Our model is explicitly low-dimensional but we believe it may be a starting point for a more general approach to extended systems. 
The transition between silent (but excitable) and spontaneous activity has long been studied in neural cultures (e.g. \cite{Beggs2003,Mazzoni2007,Benayoun2010}), and the transition may be accompanied by large scale activity that has been characterised as ``avalanches'' with a 1/f-like power spectrum. This is well-known as an outcome for dynamical systems that are driven very slowly close to a threshold, named self-organised criticality \cite{Bak1987}, and also for extended disordered nonlinear systems under periodic driving \cite{Coppersmith1987,Tang1987}. Such states are highly correlated in space and time, but not fully periodic. Such systems are based directly on generalisations of the low dimensional model we discuss here, though extensions are not simple.

Close to the bifurcation, our model could potentially be applicable to a wide variety of other systems, such as visual perception bistability \cite{Rodriguez2018} and auditory perception bistability \cite{Rankin2015, Curtu2019}, with suitable choices of bifurcation parameters. Scaling ideas similar to our discussion have also been explored in the context of multistable perception \cite{Cao2016}. Although direct access to the limit cycle dynamics of such systems might not be readily available, the processing of stimuli is often inherently oscillatory (i.e. neural activity \cite{Stiefel2016}), thus driving the choice to explore limit cycles. Alternatively, in perception research, the stimulus itself can be oscillatory \cite{Liu2012}. Bifurcations of limit cycles and bistable regimes also occur in queuing theory \cite{Collera2022}, among others.

The choice of the type of bifurcation other than the pitchfork that can be coupled to an oscillator would be based on the phenomenological knowledge of each particular system. Same also applies to the type of noise that should be used in the model, as well as it's spectrum. The exploration of these ideas, in addition to increasing the dimensionality of these models is reserved for future work.  

\section*{Acknowledgments}
The authors would like to thank Cheyne Weis and Ed Awh for helpful discussions.

\nocite{*}
\bibliography{apssamp}

\providecommand{\noopsort}[1]{}\providecommand{\singleletter}[1]{#1}%
\begin{thebibliography}{50}%
\makeatletter
\providecommand \@ifxundefined [1]{%
 \@ifx{#1\undefined}
}%
\providecommand \@ifnum [1]{%
 \ifnum #1\expandafter \@firstoftwo
 \else \expandafter \@secondoftwo
 \fi
}%
\providecommand \@ifx [1]{%
 \ifx #1\expandafter \@firstoftwo
 \else \expandafter \@secondoftwo
 \fi
}%
\providecommand \natexlab [1]{#1}%
\providecommand \enquote  [1]{``#1''}%
\providecommand \bibnamefont  [1]{#1}%
\providecommand \bibfnamefont [1]{#1}%
\providecommand \citenamefont [1]{#1}%
\providecommand \href@noop [0]{\@secondoftwo}%
\providecommand \href [0]{\begingroup \@sanitize@url \@href}%
\providecommand \@href[1]{\@@startlink{#1}\@@href}%
\providecommand \@@href[1]{\endgroup#1\@@endlink}%
\providecommand \@sanitize@url [0]{\catcode `\\12\catcode `\$12\catcode
  `\&12\catcode `\#12\catcode `\^12\catcode `\_12\catcode `\%12\relax}%
\providecommand \@@startlink[1]{}%
\providecommand \@@endlink[0]{}%
\providecommand \url  [0]{\begingroup\@sanitize@url \@url }%
\providecommand \@url [1]{\endgroup\@href {#1}{\urlprefix }}%
\providecommand \urlprefix  [0]{URL }%
\providecommand \Eprint [0]{\href }%
\providecommand \doibase [0]{https://doi.org/}%
\providecommand \selectlanguage [0]{\@gobble}%
\providecommand \bibinfo  [0]{\@secondoftwo}%
\providecommand \bibfield  [0]{\@secondoftwo}%
\providecommand \translation [1]{[#1]}%
\providecommand \BibitemOpen [0]{}%
\providecommand \bibitemStop [0]{}%
\providecommand \bibitemNoStop [0]{.\EOS\space}%
\providecommand \EOS [0]{\spacefactor3000\relax}%
\providecommand \BibitemShut  [1]{\csname bibitem#1\endcsname}%
\let\auto@bib@innerbib\@empty
\bibitem [{\citenamefont {Strogatz}(2015)}]{Strogatzbook}%
  \BibitemOpen
  \bibfield  {author} {\bibinfo {author} {\bibfnamefont {S.~H.}\ \bibnamefont
  {Strogatz}},\ }\href@noop {} {\emph {\bibinfo {title} {Nonlinear Dynamics and
  Chaos: With applications to physics, biology, chemistry, and engineering}}}\
  (\bibinfo  {publisher} {CRC Press},\ \bibinfo {year} {2015})\BibitemShut
  {NoStop}%
\bibitem [{\citenamefont {Guckenheimer}\ and\ \citenamefont
  {Holmes}(2002)}]{Guckenheimerbook}%
  \BibitemOpen
  \bibfield  {author} {\bibinfo {author} {\bibfnamefont {J.}~\bibnamefont
  {Guckenheimer}}\ and\ \bibinfo {author} {\bibfnamefont {P.}~\bibnamefont
  {Holmes}},\ }\href@noop {} {\emph {\bibinfo {title} {Nonlinear oscillations,
  dynamical systems, and bifurcations of Vector Fields}}}\ (\bibinfo
  {publisher} {Springer},\ \bibinfo {year} {2002})\BibitemShut {NoStop}%
\bibitem [{\citenamefont {Kuramoto}(2003)}]{Kuramotobook}%
  \BibitemOpen
  \bibfield  {author} {\bibinfo {author} {\bibfnamefont {Y.}~\bibnamefont
  {Kuramoto}},\ }\href@noop {} {\emph {\bibinfo {title} {Chemical oscillations,
  waves, and turbulence}}}\ (\bibinfo  {publisher} {Dover Publ},\ \bibinfo
  {year} {2003})\BibitemShut {NoStop}%
\bibitem [{\citenamefont {Dicke}(1954)}]{Dicke1954}%
  \BibitemOpen
  \bibfield  {author} {\bibinfo {author} {\bibfnamefont {R.~H.}\ \bibnamefont
  {Dicke}},\ }\bibfield  {title} {\bibinfo {title} {\textit{Coherence in
  Spontaneous Radiation Processes}},\ }\href
  {https://doi.org/10.1103/PhysRev.93.99} {\bibfield  {journal} {\bibinfo
  {journal} {Phys. Rev.}\ }\textbf {\bibinfo {volume} {93}},\ \bibinfo {pages}
  {99} (\bibinfo {year} {1954})}\BibitemShut {NoStop}%
\bibitem [{\citenamefont {Rice}\ and\ \citenamefont
  {Carmichael}(1994)}]{Rice1994}%
  \BibitemOpen
  \bibfield  {author} {\bibinfo {author} {\bibfnamefont {P.~R.}\ \bibnamefont
  {Rice}}\ and\ \bibinfo {author} {\bibfnamefont {H.~J.}\ \bibnamefont
  {Carmichael}},\ }\bibfield  {title} {\bibinfo {title} {\textit{Photon
  statistics of a cavity-QED laser: A comment on the laser--phase-transition
  analogy}},\ }\href {https://doi.org/10.1103/PhysRevA.50.4318} {\bibfield
  {journal} {\bibinfo  {journal} {Phys. Rev. A}\ }\textbf {\bibinfo {volume}
  {50}},\ \bibinfo {pages} {4318} (\bibinfo {year} {1994})}\BibitemShut
  {NoStop}%
\bibitem [{\citenamefont {Eastham}\ and\ \citenamefont
  {Littlewood}(2001)}]{Eastham2001}%
  \BibitemOpen
  \bibfield  {author} {\bibinfo {author} {\bibfnamefont {P.~R.}\ \bibnamefont
  {Eastham}}\ and\ \bibinfo {author} {\bibfnamefont {P.~B.}\ \bibnamefont
  {Littlewood}},\ }\bibfield  {title} {\bibinfo {title} {\textit{Bose
  condensation of cavity polaritons beyond the linear regime: The thermal
  equilibrium of a model microcavity}},\ }\href
  {https://doi.org/10.1103/PhysRevB.64.235101} {\bibfield  {journal} {\bibinfo
  {journal} {Phys. Rev. B}\ }\textbf {\bibinfo {volume} {64}},\ \bibinfo
  {pages} {235101} (\bibinfo {year} {2001})}\BibitemShut {NoStop}%
\bibitem [{\citenamefont {Kirton}\ and\ \citenamefont
  {Keeling}(2018)}]{Kirton2018}%
  \BibitemOpen
  \bibfield  {author} {\bibinfo {author} {\bibfnamefont {P.}~\bibnamefont
  {Kirton}}\ and\ \bibinfo {author} {\bibfnamefont {J.}~\bibnamefont
  {Keeling}},\ }\bibfield  {title} {\bibinfo {title} {\textit{Superradiant and
  lasing states in driven-dissipative Dicke models}},\ }\href
  {https://doi.org/10.1088/1367-2630/aaa11d} {\bibfield  {journal} {\bibinfo
  {journal} {New Journal of Physics}\ }\textbf {\bibinfo {volume} {20}},\
  \bibinfo {pages} {015009} (\bibinfo {year} {2018})}\BibitemShut {NoStop}%
\bibitem [{\citenamefont {Dukelsky}\ \emph {et~al.}(2004)\citenamefont
  {Dukelsky}, \citenamefont {Pittel},\ and\ \citenamefont
  {Sierra}}]{Dukelsky2004}%
  \BibitemOpen
  \bibfield  {author} {\bibinfo {author} {\bibfnamefont {J.}~\bibnamefont
  {Dukelsky}}, \bibinfo {author} {\bibfnamefont {S.}~\bibnamefont {Pittel}},\
  and\ \bibinfo {author} {\bibfnamefont {G.}~\bibnamefont {Sierra}},\
  }\bibfield  {title} {\bibinfo {title} {\textit{Colloquium: Exactly solvable
  Richardson-Gaudin models for many-body quantum systems}},\ }\href
  {https://doi.org/10.1103/RevModPhys.76.643} {\bibfield  {journal} {\bibinfo
  {journal} {Rev. Mod. Phys.}\ }\textbf {\bibinfo {volume} {76}},\ \bibinfo
  {pages} {643} (\bibinfo {year} {2004})}\BibitemShut {NoStop}%
\bibitem [{\citenamefont {Winfree}(2001)}]{winfree2001}%
  \BibitemOpen
  \bibfield  {author} {\bibinfo {author} {\bibfnamefont {A.~T.}\ \bibnamefont
  {Winfree}},\ }\href
  {https://doi.org/https://doi.org/10.1007/978-1-4757-3484-3} {\emph {\bibinfo
  {title} {The Geometry of Biological Time}}}\ (\bibinfo  {publisher} {Springer
  New York, NY},\ \bibinfo {year} {2001})\BibitemShut {NoStop}%
\bibitem [{\citenamefont {Camalet}\ \emph {et~al.}(2000)\citenamefont
  {Camalet}, \citenamefont {Duke}, \citenamefont {Jülicher},\ and\
  \citenamefont {Prost}}]{Camelet2000}%
  \BibitemOpen
  \bibfield  {author} {\bibinfo {author} {\bibfnamefont {S.}~\bibnamefont
  {Camalet}}, \bibinfo {author} {\bibfnamefont {T.}~\bibnamefont {Duke}},
  \bibinfo {author} {\bibfnamefont {F.}~\bibnamefont {Jülicher}},\ and\
  \bibinfo {author} {\bibfnamefont {J.}~\bibnamefont {Prost}},\ }\bibfield
  {title} {\bibinfo {title} {\textit{Auditory sensitivity provided by
  self-tuned critical oscillations of hair cells.}},\ }\href
  {https://doi.org/10.1073/pnas.97.7.3183} {\bibfield  {journal} {\bibinfo
  {journal} {Proc. Natl. Acad. Sci.}\ }\textbf {\bibinfo {volume} {97}},\
  \bibinfo {pages} {3183} (\bibinfo {year} {2000})}\BibitemShut {NoStop}%
\bibitem [{\citenamefont {Smith}\ \emph {et~al.}(1991)\citenamefont {Smith},
  \citenamefont {Ellenberger}, \citenamefont {Ballanyi}, \citenamefont
  {Richter},\ and\ \citenamefont {Feldman}}]{Smith1991}%
  \BibitemOpen
  \bibfield  {author} {\bibinfo {author} {\bibfnamefont {J.~C.}\ \bibnamefont
  {Smith}}, \bibinfo {author} {\bibfnamefont {H.~H.}\ \bibnamefont
  {Ellenberger}}, \bibinfo {author} {\bibfnamefont {K.}~\bibnamefont
  {Ballanyi}}, \bibinfo {author} {\bibfnamefont {D.~W.}\ \bibnamefont
  {Richter}},\ and\ \bibinfo {author} {\bibfnamefont {J.~L.}\ \bibnamefont
  {Feldman}},\ }\bibfield  {title} {\bibinfo {title} {\textit{Pre-Bötzinger
  complex: a brainstem region that may generate respiratory rhythm in
  mammals}},\ }\href {https://doi.org/doi:10.1126/science.1683005} {\bibfield
  {journal} {\bibinfo  {journal} {Science}\ }\textbf {\bibinfo {volume}
  {254}},\ \bibinfo {pages} {726} (\bibinfo {year} {1991})}\BibitemShut
  {NoStop}%
\bibitem [{\citenamefont {Rijo-Ferreira}\ and\ \citenamefont
  {Takahashi}(2019)}]{Rijo-Ferreira2019}%
  \BibitemOpen
  \bibfield  {author} {\bibinfo {author} {\bibfnamefont {F.}~\bibnamefont
  {Rijo-Ferreira}}\ and\ \bibinfo {author} {\bibfnamefont {J.~S.}\ \bibnamefont
  {Takahashi}},\ }\bibfield  {title} {\bibinfo {title} {\textit{Genomics of
  circadian rhythms in health and disease}},\ }\href
  {https://doi.org/10.1186/s13073-019-0704-0} {\bibfield  {journal} {\bibinfo
  {journal} {Genome Med.}\ }\textbf {\bibinfo {volume} {11}},\ \bibinfo {pages}
  {82} (\bibinfo {year} {2019})}\BibitemShut {NoStop}%
\bibitem [{\citenamefont {Himona}\ \emph {et~al.}(2022)\citenamefont {Himona},
  \citenamefont {Kovanis},\ and\ \citenamefont {Kominis}}]{Himona2022}%
  \BibitemOpen
  \bibfield  {author} {\bibinfo {author} {\bibfnamefont {G.}~\bibnamefont
  {Himona}}, \bibinfo {author} {\bibfnamefont {V.}~\bibnamefont {Kovanis}},\
  and\ \bibinfo {author} {\bibfnamefont {Y.}~\bibnamefont {Kominis}},\
  }\bibfield  {title} {\bibinfo {title} {\textit{Isochrons, phase response and
  synchronization dynamics of tunable photonic oscillators}},\ }\bibfield
  {journal} {\bibinfo  {journal} {Phys. Rev. Res.}\ }\textbf {\bibinfo {volume}
  {4}},\ \href {https://doi.org/10.1103/physrevresearch.4.l012039}
  {10.1103/physrevresearch.4.l012039} (\bibinfo {year} {2022})\BibitemShut
  {NoStop}%
\bibitem [{\citenamefont {Valagiannopoulos}\ and\ \citenamefont
  {Kovanis}(2021)}]{Vassilios2021}%
  \BibitemOpen
  \bibfield  {author} {\bibinfo {author} {\bibfnamefont {C.}~\bibnamefont
  {Valagiannopoulos}}\ and\ \bibinfo {author} {\bibfnamefont {V.}~\bibnamefont
  {Kovanis}},\ }\bibfield  {title} {\bibinfo {title} {\textit{Injection-Locked
  Photonic Oscillators: Legacy results and future applications}},\ }\href
  {https://doi.org/10.1109/MAP.2020.3021391} {\bibfield  {journal} {\bibinfo
  {journal} {IEEE Antennas Propag. Mag.}\ }\textbf {\bibinfo {volume} {63}},\
  \bibinfo {pages} {51} (\bibinfo {year} {2021})}\BibitemShut {NoStop}%
\bibitem [{\citenamefont {Hanai}\ \emph {et~al.}(2019)\citenamefont {Hanai},
  \citenamefont {Edelman}, \citenamefont {Ohashi},\ and\ \citenamefont
  {Littlewood}}]{Hanai2019}%
  \BibitemOpen
  \bibfield  {author} {\bibinfo {author} {\bibfnamefont {R.}~\bibnamefont
  {Hanai}}, \bibinfo {author} {\bibfnamefont {A.}~\bibnamefont {Edelman}},
  \bibinfo {author} {\bibfnamefont {Y.}~\bibnamefont {Ohashi}},\ and\ \bibinfo
  {author} {\bibfnamefont {P.~B.}\ \bibnamefont {Littlewood}},\ }\bibfield
  {title} {\bibinfo {title} {\textit{Non-Hermitian Phase Transition from a
  Polariton Bose-Einstein Condensate to a Photon Laser}},\ }\href
  {https://doi.org/10.1103/PhysRevLett.122.185301} {\bibfield  {journal}
  {\bibinfo  {journal} {Phys. Rev. Lett.}\ }\textbf {\bibinfo {volume} {122}},\
  \bibinfo {pages} {185301} (\bibinfo {year} {2019})}\BibitemShut {NoStop}%
\bibitem [{\citenamefont {Fruchart}\ \emph {et~al.}(2021)\citenamefont
  {Fruchart}, \citenamefont {Hanai}, \citenamefont {Littlewood},\ and\
  \citenamefont {Vitelli}}]{Fruchart2021}%
  \BibitemOpen
  \bibfield  {author} {\bibinfo {author} {\bibfnamefont {M.}~\bibnamefont
  {Fruchart}}, \bibinfo {author} {\bibfnamefont {R.}~\bibnamefont {Hanai}},
  \bibinfo {author} {\bibfnamefont {P.~B.}\ \bibnamefont {Littlewood}},\ and\
  \bibinfo {author} {\bibfnamefont {V.}~\bibnamefont {Vitelli}},\ }\bibfield
  {title} {\bibinfo {title} {\textit{Non-reciprocal phase transitions}},\
  }\href {https://doi.org/10.1038/s41586-021-03375-9} {\bibfield  {journal}
  {\bibinfo  {journal} {Nature}\ }\textbf {\bibinfo {volume} {592}},\ \bibinfo
  {pages} {363} (\bibinfo {year} {2021})}\BibitemShut {NoStop}%
\bibitem [{\citenamefont {McNamara}\ and\ \citenamefont
  {Wiesenfeld}(1989)}]{Mcnamara}%
  \BibitemOpen
  \bibfield  {author} {\bibinfo {author} {\bibfnamefont {B.}~\bibnamefont
  {McNamara}}\ and\ \bibinfo {author} {\bibfnamefont {K.}~\bibnamefont
  {Wiesenfeld}},\ }\bibfield  {title} {\bibinfo {title} {\textit{Theory of
  Stochastic Resonance}},\ }\href@noop {} {\bibfield  {journal} {\bibinfo
  {journal} {Phys. Rev. A}\ }\textbf {\bibinfo {volume} {39}},\ \bibinfo
  {pages} {4854} (\bibinfo {year} {1989})}\BibitemShut {NoStop}%
\bibitem [{\citenamefont {Gammaitoni}\ \emph {et~al.}(1998)\citenamefont
  {Gammaitoni}, \citenamefont {Hänggi}, \citenamefont {Jung},\ and\
  \citenamefont {Marchesoni}}]{SR2}%
  \BibitemOpen
  \bibfield  {author} {\bibinfo {author} {\bibfnamefont {L.}~\bibnamefont
  {Gammaitoni}}, \bibinfo {author} {\bibfnamefont {P.}~\bibnamefont {Hänggi}},
  \bibinfo {author} {\bibfnamefont {P.}~\bibnamefont {Jung}},\ and\ \bibinfo
  {author} {\bibfnamefont {F.}~\bibnamefont {Marchesoni}},\ }\bibfield  {title}
  {\bibinfo {title} {\textit{Stochastic resonance}},\ }\href
  {https://doi.org/10.1103/revmodphys.70.223} {\bibfield  {journal} {\bibinfo
  {journal} {Rev. Mod. Phys.}\ }\textbf {\bibinfo {volume} {70}},\ \bibinfo
  {pages} {223–287} (\bibinfo {year} {1998})}\BibitemShut {NoStop}%
\bibitem [{\citenamefont {{Weis}}\ \emph {et~al.}(2022)\citenamefont {{Weis}},
  \citenamefont {{Fruchart}}, \citenamefont {{Hanai}}, \citenamefont
  {{Kawagoe}}, \citenamefont {{Littlewood}},\ and\ \citenamefont
  {{Vitelli}}}]{Weis}%
  \BibitemOpen
  \bibfield  {author} {\bibinfo {author} {\bibfnamefont {C.}~\bibnamefont
  {{Weis}}}, \bibinfo {author} {\bibfnamefont {M.}~\bibnamefont {{Fruchart}}},
  \bibinfo {author} {\bibfnamefont {R.}~\bibnamefont {{Hanai}}}, \bibinfo
  {author} {\bibfnamefont {K.}~\bibnamefont {{Kawagoe}}}, \bibinfo {author}
  {\bibfnamefont {P.~B.}\ \bibnamefont {{Littlewood}}},\ and\ \bibinfo {author}
  {\bibfnamefont {V.}~\bibnamefont {{Vitelli}}},\ }\bibfield  {title} {\bibinfo
  {title} {{\textit{Coalescence of Attractors: Exceptional points in non-linear
  dynamical systems}}},\ }\href@noop {} {\bibfield  {journal} {\bibinfo
  {journal} {arXiv e-prints}\ ,\ \bibinfo {eid} {arXiv:2207.11667}} (\bibinfo
  {year} {2022})}\BibitemShut {NoStop}%
\bibitem [{\citenamefont {Goldobin}\ \emph {et~al.}(2010)\citenamefont
  {Goldobin}, \citenamefont {Teramae}, \citenamefont {Nakao},\ and\
  \citenamefont {Ermentrout}}]{LimitCycleNoise}%
  \BibitemOpen
  \bibfield  {author} {\bibinfo {author} {\bibfnamefont {D.~S.}\ \bibnamefont
  {Goldobin}}, \bibinfo {author} {\bibfnamefont {J.-n.}\ \bibnamefont
  {Teramae}}, \bibinfo {author} {\bibfnamefont {H.}~\bibnamefont {Nakao}},\
  and\ \bibinfo {author} {\bibfnamefont {G.~B.}\ \bibnamefont {Ermentrout}},\
  }\bibfield  {title} {\bibinfo {title} {\textit{Dynamics of Limit-Cycle
  Oscillators Subject to General Noise}},\ }\href
  {https://doi.org/10.1103/PhysRevLett.105.154101} {\bibfield  {journal}
  {\bibinfo  {journal} {Phys. Rev. Lett.}\ }\textbf {\bibinfo {volume} {105}},\
  \bibinfo {pages} {154101} (\bibinfo {year} {2010})}\BibitemShut {NoStop}%
\bibitem [{\citenamefont {Rosenblum}\ and\ \citenamefont
  {Pikovsky}(2004)}]{Example1}%
  \BibitemOpen
  \bibfield  {author} {\bibinfo {author} {\bibfnamefont {M.~G.}\ \bibnamefont
  {Rosenblum}}\ and\ \bibinfo {author} {\bibfnamefont {A.~S.}\ \bibnamefont
  {Pikovsky}},\ }\bibfield  {title} {\bibinfo {title} {\textit{Controlling
  Synchronization in an Ensemble of Globally Coupled Oscillator}s},\ }\href
  {https://doi.org/10.1103/PhysRevLett.92.114102} {\bibfield  {journal}
  {\bibinfo  {journal} {Phys. Rev. Lett.}\ }\textbf {\bibinfo {volume} {92}},\
  \bibinfo {pages} {114102} (\bibinfo {year} {2004})}\BibitemShut {NoStop}%
\bibitem [{\citenamefont {Grenfell}\ \emph {et~al.}(1998)\citenamefont
  {Grenfell}, \citenamefont {Wilson}, \citenamefont {Finkenstädt},
  \citenamefont {Coulson}, \citenamefont {Murray}, \citenamefont {Albon},
  \citenamefont {Pemberton}, \citenamefont {Clutton-Brock},\ and\ \citenamefont
  {Crawley}}]{Example2}%
  \BibitemOpen
  \bibfield  {author} {\bibinfo {author} {\bibfnamefont {B.~T.}\ \bibnamefont
  {Grenfell}}, \bibinfo {author} {\bibfnamefont {K.}~\bibnamefont {Wilson}},
  \bibinfo {author} {\bibfnamefont {B.~F.}\ \bibnamefont {Finkenstädt}},
  \bibinfo {author} {\bibfnamefont {T.~N.}\ \bibnamefont {Coulson}}, \bibinfo
  {author} {\bibfnamefont {S.}~\bibnamefont {Murray}}, \bibinfo {author}
  {\bibfnamefont {S.~D.}\ \bibnamefont {Albon}}, \bibinfo {author}
  {\bibfnamefont {J.~M.}\ \bibnamefont {Pemberton}}, \bibinfo {author}
  {\bibfnamefont {T.~H.}\ \bibnamefont {Clutton-Brock}},\ and\ \bibinfo
  {author} {\bibfnamefont {M.~J.}\ \bibnamefont {Crawley}},\ }\bibfield
  {title} {\bibinfo {title} {\textit{Noise and determinism in Synchronized
  Sheep Dynamics}},\ }\href {https://doi.org/10.1038/29291} {\bibfield
  {journal} {\bibinfo  {journal} {Nature}\ }\textbf {\bibinfo {volume} {394}},\
  \bibinfo {pages} {674–677} (\bibinfo {year} {1998})}\BibitemShut {NoStop}%
\bibitem [{\citenamefont {Yoshimura}\ and\ \citenamefont
  {Arai}(2008)}]{Example3}%
  \BibitemOpen
  \bibfield  {author} {\bibinfo {author} {\bibfnamefont {K.}~\bibnamefont
  {Yoshimura}}\ and\ \bibinfo {author} {\bibfnamefont {K.}~\bibnamefont
  {Arai}},\ }\bibfield  {title} {\bibinfo {title} {\textit{Phase Reduction of
  Stochastic Limit Cycle Oscillators}},\ }\href
  {https://doi.org/10.1103/PhysRevLett.101.154101} {\bibfield  {journal}
  {\bibinfo  {journal} {Phys. Rev. Lett.}\ }\textbf {\bibinfo {volume} {101}},\
  \bibinfo {pages} {154101} (\bibinfo {year} {2008})}\BibitemShut {NoStop}%
\bibitem [{\citenamefont {Hertz}\ \emph {et~al.}(2016)\citenamefont {Hertz},
  \citenamefont {Roudi},\ and\ \citenamefont {Sollich}}]{Hertz}%
  \BibitemOpen
  \bibfield  {author} {\bibinfo {author} {\bibfnamefont {J.}~\bibnamefont
  {Hertz}}, \bibinfo {author} {\bibfnamefont {Y.}~\bibnamefont {Roudi}},\ and\
  \bibinfo {author} {\bibfnamefont {P.}~\bibnamefont {Sollich}},\ }\bibfield
  {title} {\bibinfo {title} {\textit{Path Integral Methods for the Dynamics of
  Stochastic and Disordered Systems}},\ }\href@noop {} {\bibfield  {journal}
  {\bibinfo  {journal} {J. Phys. A: Math. Theor.}\ }\textbf {\bibinfo {volume}
  {50}},\ \bibinfo {pages} {033001} (\bibinfo {year} {2016})}\BibitemShut
  {NoStop}%
\bibitem [{\citenamefont {Gardiner}(1997)}]{Gardiner}%
  \BibitemOpen
  \bibfield  {author} {\bibinfo {author} {\bibfnamefont {C.~W.}\ \bibnamefont
  {Gardiner}},\ }\href@noop {} {\emph {\bibinfo {title} {Handbook of Stochastic
  Methods}}}\ (\bibinfo  {publisher} {Springer-Verlag},\ \bibinfo {year}
  {1997})\BibitemShut {NoStop}%
\bibitem [{\citenamefont {Martin}\ \emph {et~al.}(1973)\citenamefont {Martin},
  \citenamefont {Siggia},\ and\ \citenamefont {Rose}}]{MSR1}%
  \BibitemOpen
  \bibfield  {author} {\bibinfo {author} {\bibfnamefont {P.~C.}\ \bibnamefont
  {Martin}}, \bibinfo {author} {\bibfnamefont {E.~D.}\ \bibnamefont {Siggia}},\
  and\ \bibinfo {author} {\bibfnamefont {H.~A.}\ \bibnamefont {Rose}},\
  }\bibfield  {title} {\bibinfo {title} {\textit{Statistical Dynamics of
  Classical Systems}},\ }\href {https://doi.org/10.1103/physreva.8.423}
  {\bibfield  {journal} {\bibinfo  {journal} {Phys. Rev. A}\ }\textbf {\bibinfo
  {volume} {8}},\ \bibinfo {pages} {423–437} (\bibinfo {year}
  {1973})}\BibitemShut {NoStop}%
\bibitem [{\citenamefont {{De Dominicis}}\ and\ \citenamefont
  {Peliti}(1978)}]{MSR2}%
  \BibitemOpen
  \bibfield  {author} {\bibinfo {author} {\bibfnamefont {C.}~\bibnamefont {{De
  Dominicis}}}\ and\ \bibinfo {author} {\bibfnamefont {L.}~\bibnamefont
  {Peliti}},\ }\bibfield  {title} {\bibinfo {title} {\textit{Field-theory
  renormalization and critical dynamics above $T_c$: Helium, antiferromagnets,
  and liquid-gas systems}},\ }\href@noop {} {\bibfield  {journal} {\bibinfo
  {journal} {Phys. Rev. B}\ }\textbf {\bibinfo {volume} {18}},\ \bibinfo
  {pages} {353} (\bibinfo {year} {1978})}\BibitemShut {NoStop}%
\bibitem [{\citenamefont {Atland}\ and\ \citenamefont {Simons}(2010)}]{Atland}%
  \BibitemOpen
  \bibfield  {author} {\bibinfo {author} {\bibfnamefont {A.}~\bibnamefont
  {Atland}}\ and\ \bibinfo {author} {\bibfnamefont {B.}~\bibnamefont
  {Simons}},\ }\href@noop {} {\emph {\bibinfo {title} {Condensed Matter Field
  Theory}}}\ (\bibinfo  {publisher} {Cambridge University Press},\ \bibinfo
  {year} {2010})\BibitemShut {NoStop}%
\bibitem [{\citenamefont {Kamenev}(2011)}]{Kamenev}%
  \BibitemOpen
  \bibfield  {author} {\bibinfo {author} {\bibfnamefont {A.}~\bibnamefont
  {Kamenev}},\ }\href@noop {} {\emph {\bibinfo {title} {Field Theory of
  Non-Equilibrium Systems}}}\ (\bibinfo  {publisher} {Cambridge University
  Press},\ \bibinfo {year} {2011})\BibitemShut {NoStop}%
\bibitem [{\citenamefont {Faber}\ and\ \citenamefont
  {Bozovic}(2019{\natexlab{a}})}]{HopfExample1}%
  \BibitemOpen
  \bibfield  {author} {\bibinfo {author} {\bibfnamefont {J.}~\bibnamefont
  {Faber}}\ and\ \bibinfo {author} {\bibfnamefont {D.}~\bibnamefont
  {Bozovic}},\ }\bibfield  {title} {\bibinfo {title} {\textit{Noise-induced
  chaos and signal detection by the nonisochronous Hopf oscillator}},\ }\href
  {https://doi.org/10.1063/1.5091938} {\bibfield  {journal} {\bibinfo
  {journal} {Chaos}\ }\textbf {\bibinfo {volume} {29}},\ \bibinfo {pages}
  {043132} (\bibinfo {year} {2019}{\natexlab{a}})}\BibitemShut {NoStop}%
\bibitem [{\citenamefont {Juel}\ \emph {et~al.}(1997)\citenamefont {Juel},
  \citenamefont {Darbyshire},\ and\ \citenamefont {Mullin}}]{HopfExample2}%
  \BibitemOpen
  \bibfield  {author} {\bibinfo {author} {\bibfnamefont {A.}~\bibnamefont
  {Juel}}, \bibinfo {author} {\bibfnamefont {A.~G.}\ \bibnamefont
  {Darbyshire}},\ and\ \bibinfo {author} {\bibfnamefont {T.}~\bibnamefont
  {Mullin}},\ }\bibfield  {title} {\bibinfo {title} {\textit{The effect of
  noise on Pitchfork and Hopf bifurcations}},\ }\href
  {https://doi.org/10.1098/rspa.1997.0140} {\bibfield  {journal} {\bibinfo
  {journal} {Proc. R. Soc. A}\ }\textbf {\bibinfo {volume} {453}},\ \bibinfo
  {pages} {2627–2647} (\bibinfo {year} {1997})}\BibitemShut {NoStop}%
\bibitem [{\citenamefont {Doan}\ \emph {et~al.}(2018)\citenamefont {Doan},
  \citenamefont {Engel}, \citenamefont {Lamb},\ and\ \citenamefont
  {Rasmussen}}]{HopfExample3}%
  \BibitemOpen
  \bibfield  {author} {\bibinfo {author} {\bibfnamefont {T.~S.}\ \bibnamefont
  {Doan}}, \bibinfo {author} {\bibfnamefont {M.}~\bibnamefont {Engel}},
  \bibinfo {author} {\bibfnamefont {J.~S.}\ \bibnamefont {Lamb}},\ and\
  \bibinfo {author} {\bibfnamefont {M.}~\bibnamefont {Rasmussen}},\ }\bibfield
  {title} {\bibinfo {title} {\textit{Hopf bifurcation with additive noise}},\
  }\href {https://doi.org/10.1088/1361-6544/aad208} {\bibfield  {journal}
  {\bibinfo  {journal} {Nonlinearity}\ }\textbf {\bibinfo {volume} {31}},\
  \bibinfo {pages} {4567–4601} (\bibinfo {year} {2018})}\BibitemShut
  {NoStop}%
\bibitem [{\citenamefont {Risler}\ \emph {et~al.}(2004)\citenamefont {Risler},
  \citenamefont {Prost},\ and\ \citenamefont {Jülicher}}]{HopfExample4}%
  \BibitemOpen
  \bibfield  {author} {\bibinfo {author} {\bibfnamefont {T.}~\bibnamefont
  {Risler}}, \bibinfo {author} {\bibfnamefont {J.}~\bibnamefont {Prost}},\ and\
  \bibinfo {author} {\bibfnamefont {F.}~\bibnamefont {Jülicher}},\ }\bibfield
  {title} {\bibinfo {title} {\textit{Universal critical behavior of noisy
  coupled oscillators}},\ }\href
  {https://doi.org/10.1103/physrevlett.93.175702} {\bibfield  {journal}
  {\bibinfo  {journal} {Phys. Rev. Lett.}\ }\textbf {\bibinfo {volume} {93}},\
  \bibinfo {pages} {175702} (\bibinfo {year} {2004})}\BibitemShut {NoStop}%
\bibitem [{\citenamefont {Faber}\ and\ \citenamefont
  {Bozovic}(2019{\natexlab{b}})}]{HearingHopf}%
  \BibitemOpen
  \bibfield  {author} {\bibinfo {author} {\bibfnamefont {J.}~\bibnamefont
  {Faber}}\ and\ \bibinfo {author} {\bibfnamefont {D.}~\bibnamefont
  {Bozovic}},\ }\bibfield  {title} {\bibinfo {title} {\textit{Chaotic dynamics
  enhance the sensitivity of inner ear hair cells}},\ }\href
  {https://doi.org/10.1038/s41598-019-54952-y} {\bibfield  {journal} {\bibinfo
  {journal} {Sci. Rep.}\ }\textbf {\bibinfo {volume} {9}},\ \bibinfo {pages}
  {18394} (\bibinfo {year} {2019}{\natexlab{b}})}\BibitemShut {NoStop}%
\bibitem [{\citenamefont {Chow}\ and\ \citenamefont {Buice}(2015)}]{Chow}%
  \BibitemOpen
  \bibfield  {author} {\bibinfo {author} {\bibfnamefont {C.}~\bibnamefont
  {Chow}}\ and\ \bibinfo {author} {\bibfnamefont {M.}~\bibnamefont {Buice}},\
  }\bibfield  {title} {\bibinfo {title} {\textit{Path Integral Methods for
  Stochastic Differential Equations}},\ }\href@noop {} {\bibfield  {journal}
  {\bibinfo  {journal} {J. Math. Neurosc.}\ }\textbf {\bibinfo {volume} {5}}
  (\bibinfo {year} {2015})}\BibitemShut {NoStop}%
\bibitem [{\citenamefont {Barabási}\ and\ \citenamefont
  {Stanley}(1995)}]{Barabasi}%
  \BibitemOpen
  \bibfield  {author} {\bibinfo {author} {\bibfnamefont {A.-L.}\ \bibnamefont
  {Barabási}}\ and\ \bibinfo {author} {\bibfnamefont {H.~E.}\ \bibnamefont
  {Stanley}},\ }\href@noop {} {\emph {\bibinfo {title} {Fractal Concepts in
  Surface Growth}}}\ (\bibinfo  {publisher} {Cambridge University Press},\
  \bibinfo {year} {1995})\BibitemShut {NoStop}%
\bibitem [{\citenamefont {Binney}\ \emph {et~al.}(1992)\citenamefont {Binney},
  \citenamefont {Dowrick}, \citenamefont {Fisher},\ and\ \citenamefont
  {Newman}}]{Binney}%
  \BibitemOpen
  \bibfield  {author} {\bibinfo {author} {\bibfnamefont {J.}~\bibnamefont
  {Binney}}, \bibinfo {author} {\bibfnamefont {N.}~\bibnamefont {Dowrick}},
  \bibinfo {author} {\bibfnamefont {A.}~\bibnamefont {Fisher}},\ and\ \bibinfo
  {author} {\bibfnamefont {M.}~\bibnamefont {Newman}},\ }\href@noop {} {\emph
  {\bibinfo {title} {The Theory of Critical Phenomena: An Introduction to the
  Renormalization Group}}}\ (\bibinfo  {publisher} {Clarendon Press, Oxford},\
  \bibinfo {year} {1992})\BibitemShut {NoStop}%
\bibitem [{\citenamefont {Beggs}\ and\ \citenamefont
  {Plenz}(2003)}]{Beggs2003}%
  \BibitemOpen
  \bibfield  {author} {\bibinfo {author} {\bibfnamefont {J.~M.}\ \bibnamefont
  {Beggs}}\ and\ \bibinfo {author} {\bibfnamefont {D.}~\bibnamefont {Plenz}},\
  }\bibfield  {title} {\bibinfo {title} {\textit{Neuronal avalanches in
  neocortical circuits}},\ }\href {https://doi.org/doi:
  10.1523/JNEUROSCI.23-35-11167.2003.} {\bibfield  {journal} {\bibinfo
  {journal} {J Neurosci.}\ }\textbf {\bibinfo {volume} {23}},\ \bibinfo {pages}
  {11167} (\bibinfo {year} {2003})}\BibitemShut {NoStop}%
\bibitem [{\citenamefont {Mazzoni}\ \emph {et~al.}(2007)\citenamefont
  {Mazzoni}, \citenamefont {Broccard}, \citenamefont {Garcia-Perez},
  \citenamefont {Bonifazi}, \citenamefont {Ruaro},\ and\ \citenamefont
  {Torre}}]{Mazzoni2007}%
  \BibitemOpen
  \bibfield  {author} {\bibinfo {author} {\bibfnamefont {A.}~\bibnamefont
  {Mazzoni}}, \bibinfo {author} {\bibfnamefont {F.~D.}\ \bibnamefont
  {Broccard}}, \bibinfo {author} {\bibfnamefont {E.}~\bibnamefont
  {Garcia-Perez}}, \bibinfo {author} {\bibfnamefont {P.}~\bibnamefont
  {Bonifazi}}, \bibinfo {author} {\bibfnamefont {M.~E.}\ \bibnamefont
  {Ruaro}},\ and\ \bibinfo {author} {\bibfnamefont {V.}~\bibnamefont {Torre}},\
  }\bibfield  {title} {\bibinfo {title} {\textit{On the Dynamics of the
  Spontaneous Activity in Neuronal Networks}},\ }\href
  {https://doi.org/https://doi.org/10.1371/journal.pone.0000439} {\bibfield
  {journal} {\bibinfo  {journal} {PLoS ONE}\ }\textbf {\bibinfo {volume} {2}},\
  \bibinfo {pages} {e439} (\bibinfo {year} {2007})}\BibitemShut {NoStop}%
\bibitem [{\citenamefont {Benayoun}\ \emph {et~al.}(2010)\citenamefont
  {Benayoun}, \citenamefont {Cowan}, \citenamefont {van Drongelen},\ and\
  \citenamefont {Wallace}}]{Benayoun2010}%
  \BibitemOpen
  \bibfield  {author} {\bibinfo {author} {\bibfnamefont {M.}~\bibnamefont
  {Benayoun}}, \bibinfo {author} {\bibfnamefont {J.~D.}\ \bibnamefont {Cowan}},
  \bibinfo {author} {\bibfnamefont {W.}~\bibnamefont {van Drongelen}},\ and\
  \bibinfo {author} {\bibfnamefont {E.}~\bibnamefont {Wallace}},\ }\bibfield
  {title} {\bibinfo {title} {\textit{Avalanches in a Stochastic Model of
  Spiking Neurons}},\ }\href {https://doi.org/doi:10.1371/journal.pcbi.1000846}
  {\bibfield  {journal} {\bibinfo  {journal} {PLoS Comput Biol}\ }\textbf
  {\bibinfo {volume} {6}},\ \bibinfo {pages} {1000846} (\bibinfo {year}
  {2010})}\BibitemShut {NoStop}%
\bibitem [{\citenamefont {Bak}\ \emph {et~al.}(1987)\citenamefont {Bak},
  \citenamefont {Tang},\ and\ \citenamefont {Wiesenfeld}}]{Bak1987}%
  \BibitemOpen
  \bibfield  {author} {\bibinfo {author} {\bibfnamefont {P.}~\bibnamefont
  {Bak}}, \bibinfo {author} {\bibfnamefont {C.}~\bibnamefont {Tang}},\ and\
  \bibinfo {author} {\bibfnamefont {K.}~\bibnamefont {Wiesenfeld}},\ }\bibfield
   {title} {\bibinfo {title} {\textit{Self-organized criticality: An
  explanation of the 1/f noise}},\ }\href
  {https://doi.org/10.1103/PhysRevLett.59.381} {\bibfield  {journal} {\bibinfo
  {journal} {Phys. Rev. Lett.}\ }\textbf {\bibinfo {volume} {59}},\ \bibinfo
  {pages} {381} (\bibinfo {year} {1987})}\BibitemShut {NoStop}%
\bibitem [{\citenamefont {Coppersmith}\ and\ \citenamefont
  {Littlewood}(1987)}]{Coppersmith1987}%
  \BibitemOpen
  \bibfield  {author} {\bibinfo {author} {\bibfnamefont {S.~N.}\ \bibnamefont
  {Coppersmith}}\ and\ \bibinfo {author} {\bibfnamefont {P.~B.}\ \bibnamefont
  {Littlewood}},\ }\bibfield  {title} {\bibinfo {title} {\textit{Pulse-duration
  memory effect and deformable charge-density waves}},\ }\href
  {https://doi.org/10.1103/PhysRevB.36.311} {\bibfield  {journal} {\bibinfo
  {journal} {Phys. Rev. B}\ }\textbf {\bibinfo {volume} {36}},\ \bibinfo
  {pages} {311} (\bibinfo {year} {1987})}\BibitemShut {NoStop}%
\bibitem [{\citenamefont {Tang}\ \emph {et~al.}(1987)\citenamefont {Tang},
  \citenamefont {Wiesenfeld}, \citenamefont {Bak}, \citenamefont
  {Coppersmith},\ and\ \citenamefont {Littlewood}}]{Tang1987}%
  \BibitemOpen
  \bibfield  {author} {\bibinfo {author} {\bibfnamefont {C.}~\bibnamefont
  {Tang}}, \bibinfo {author} {\bibfnamefont {K.}~\bibnamefont {Wiesenfeld}},
  \bibinfo {author} {\bibfnamefont {P.}~\bibnamefont {Bak}}, \bibinfo {author}
  {\bibfnamefont {S.}~\bibnamefont {Coppersmith}},\ and\ \bibinfo {author}
  {\bibfnamefont {P.}~\bibnamefont {Littlewood}},\ }\bibfield  {title}
  {\bibinfo {title} {\textit{Phase organization}},\ }\href
  {https://doi.org/10.1103/PhysRevLett.58.1161} {\bibfield  {journal} {\bibinfo
   {journal} {Phys. Rev. Lett.}\ }\textbf {\bibinfo {volume} {58}},\ \bibinfo
  {pages} {1161} (\bibinfo {year} {1987})}\BibitemShut {NoStop}%
\bibitem [{\citenamefont {Rodríguez~Martínez}\ and\ \citenamefont
  {Castillo~Parra}(2018)}]{Rodriguez2018}%
  \BibitemOpen
  \bibfield  {author} {\bibinfo {author} {\bibfnamefont {G.~A.}\ \bibnamefont
  {Rodríguez~Martínez}}\ and\ \bibinfo {author} {\bibfnamefont
  {H.}~\bibnamefont {Castillo~Parra}},\ }\bibfield  {title} {\bibinfo {title}
  {\textit{Bistable perception: Neural bases and usefulness in psychological
  research}},\ }\href {https://doi.org/10.21500/20112084.3375} {\bibfield
  {journal} {\bibinfo  {journal} {Int. J. Psychol. Res.}\ }\textbf {\bibinfo
  {volume} {11}},\ \bibinfo {pages} {63–76} (\bibinfo {year}
  {2018})}\BibitemShut {NoStop}%
\bibitem [{\citenamefont {Rankin}\ \emph {et~al.}(2015)\citenamefont {Rankin},
  \citenamefont {Sussman},\ and\ \citenamefont {Rinzel}}]{Rankin2015}%
  \BibitemOpen
  \bibfield  {author} {\bibinfo {author} {\bibfnamefont {J.}~\bibnamefont
  {Rankin}}, \bibinfo {author} {\bibfnamefont {E.}~\bibnamefont {Sussman}},\
  and\ \bibinfo {author} {\bibfnamefont {J.}~\bibnamefont {Rinzel}},\
  }\bibfield  {title} {\bibinfo {title} {\textit{Neuromechanistic model of
  auditory bistability}},\ }\bibfield  {journal} {\bibinfo  {journal} {PLoS
  Comput. Biol.}\ }\textbf {\bibinfo {volume} {11}},\ \href
  {https://doi.org/10.1371/journal.pcbi.1004555} {10.1371/journal.pcbi.1004555}
  (\bibinfo {year} {2015})\BibitemShut {NoStop}%
\bibitem [{\citenamefont {Curtu}\ \emph {et~al.}(2019)\citenamefont {Curtu},
  \citenamefont {Wang}, \citenamefont {Brunton},\ and\ \citenamefont
  {Nourski}}]{Curtu2019}%
  \BibitemOpen
  \bibfield  {author} {\bibinfo {author} {\bibfnamefont {R.}~\bibnamefont
  {Curtu}}, \bibinfo {author} {\bibfnamefont {X.}~\bibnamefont {Wang}},
  \bibinfo {author} {\bibfnamefont {B.~W.}\ \bibnamefont {Brunton}},\ and\
  \bibinfo {author} {\bibfnamefont {K.~V.}\ \bibnamefont {Nourski}},\
  }\bibfield  {title} {\bibinfo {title} {\textit{Neural signatures of auditory
  perceptual bistability revealed by large-scale human intracranial
  recordings}},\ }\href {https://doi.org/10.1523/jneurosci.0655-18.2019}
  {\bibfield  {journal} {\bibinfo  {journal} {J. Neurosc.}\ }\textbf {\bibinfo
  {volume} {39}},\ \bibinfo {pages} {6482–6497} (\bibinfo {year}
  {2019})}\BibitemShut {NoStop}%
\bibitem [{\citenamefont {Cao}\ \emph {et~al.}(2016)\citenamefont {Cao},
  \citenamefont {Pastukhov}, \citenamefont {Mattia},\ and\ \citenamefont
  {Braun}}]{Cao2016}%
  \BibitemOpen
  \bibfield  {author} {\bibinfo {author} {\bibfnamefont {R.}~\bibnamefont
  {Cao}}, \bibinfo {author} {\bibfnamefont {A.}~\bibnamefont {Pastukhov}},
  \bibinfo {author} {\bibfnamefont {M.}~\bibnamefont {Mattia}},\ and\ \bibinfo
  {author} {\bibfnamefont {J.}~\bibnamefont {Braun}},\ }\bibfield  {title}
  {\bibinfo {title} {\textit{Collective activity of many bistable assemblies
  reproduces characteristic dynamics of multistable perception}},\ }\href
  {https://doi.org/10.1523/jneurosci.4626-15.2016} {\bibfield  {journal}
  {\bibinfo  {journal} {J. Neurosc.}\ }\textbf {\bibinfo {volume} {36}},\
  \bibinfo {pages} {6957–6972} (\bibinfo {year} {2016})}\BibitemShut
  {NoStop}%
\bibitem [{\citenamefont {Stiefel}\ and\ \citenamefont
  {Ermentrout}(2016)}]{Stiefel2016}%
  \BibitemOpen
  \bibfield  {author} {\bibinfo {author} {\bibfnamefont {K.~M.}\ \bibnamefont
  {Stiefel}}\ and\ \bibinfo {author} {\bibfnamefont {G.~B.}\ \bibnamefont
  {Ermentrout}},\ }\bibfield  {title} {\bibinfo {title} {\textit{Neurons as
  oscillators}},\ }\href {https://doi.org/10.1152/jn.00525.2015} {\bibfield
  {journal} {\bibinfo  {journal} {J. Neurophysiol.}\ }\textbf {\bibinfo
  {volume} {116}},\ \bibinfo {pages} {2950–2960} (\bibinfo {year}
  {2016})}\BibitemShut {NoStop}%
\bibitem [{\citenamefont {Liu}\ \emph {et~al.}(2012)\citenamefont {Liu},
  \citenamefont {Tzeng}, \citenamefont {Hung}, \citenamefont {Tseng},\ and\
  \citenamefont {Juan}}]{Liu2012}%
  \BibitemOpen
  \bibfield  {author} {\bibinfo {author} {\bibfnamefont {C.-H.}\ \bibnamefont
  {Liu}}, \bibinfo {author} {\bibfnamefont {O.~J.}\ \bibnamefont {Tzeng}},
  \bibinfo {author} {\bibfnamefont {D.~L.}\ \bibnamefont {Hung}}, \bibinfo
  {author} {\bibfnamefont {P.}~\bibnamefont {Tseng}},\ and\ \bibinfo {author}
  {\bibfnamefont {C.-H.}\ \bibnamefont {Juan}},\ }\bibfield  {title} {\bibinfo
  {title} {\textit{Investigation of bistable perception with the “Silhouette
  spinner”: Sit still, spin the dancer with your will}},\ }\href
  {https://doi.org/10.1016/j.visres.2012.03.005} {\bibfield  {journal}
  {\bibinfo  {journal} {Vision Res.}\ }\textbf {\bibinfo {volume} {60}},\
  \bibinfo {pages} {34–39} (\bibinfo {year} {2012})}\BibitemShut {NoStop}%
\bibitem [{\citenamefont {Collera}(2022)}]{Collera2022}%
  \BibitemOpen
  \bibfield  {author} {\bibinfo {author} {\bibfnamefont {J.~A.}\ \bibnamefont
  {Collera}},\ }\bibfield  {title} {\bibinfo {title} {\textit{Bubbling,
  bistable limit cycles and quasi-periodic oscillations in queues with delayed
  information}},\ }\href {https://doi.org/10.3390/sym14020376} {\bibfield
  {journal} {\bibinfo  {journal} {Symmetry}\ }\textbf {\bibinfo {volume}
  {14}},\ \bibinfo {pages} {376} (\bibinfo {year} {2022})}\BibitemShut
  {NoStop}%
\end{thebibliography}%

\end{document}